\documentclass[12pt]{article}
\usepackage[mathscr]{eucal}

\usepackage{epsfig,amsfonts}
\usepackage[fleqn]{amsmath}
\usepackage{amsthm,amssymb}
\usepackage{graphicx,psfrag}
\usepackage{hhline}
\usepackage{cite}

\makeatletter
\@addtoreset{equation}{section}
\makeatother

\usepackage{amsxtra}
\usepackage{color}

\def\be{\begin{equation}}
\def\ee{\end{equation}}
\def\bdm{\begin{displaymath}}
\def\edm{\end{displaymath}}
\def\bea{\begin{eqnarray}}
\def\eea{\end{eqnarray}}

\def\sgn{{\rm sgn}}

\def\ri{{\rm i}}

\def\XXint#1#2#3{{\setbox0=\hbox{$#1{#2#3}{\int}$}
    \vcenter{\hbox{$#2#3$}}\kern-.5\wd0}}

\newcommand{\rd}{\mbox{d}}
\newcommand{\re}{\mbox{e}}

\begin{document}

\begin{titlepage}
\begin{flushright}
RUNHETC-2015-04\\
\end{flushright}

\vspace{1.3cm}

\bigskip

\begin{center}
\begin{LARGE}

{\bf Fidelities in the spin-boson model}

\end{LARGE}
\vspace{1.3cm}

\begin{large}

{\bf  Sergei  L. Lukyanov}$^{1,2}$

\end{large}

\vspace{1.cm}

${}^{1}$NHETC, Department of Physics and Astronomy\\
     Rutgers University\\
     Piscataway, NJ 08855-0849, USA\\
\vspace{.2cm}
and\\
\vspace{.2cm}
${}^{2}$L.D. Landau Institute for Theoretical Physics\\
  Chernogolovka, 142432, Russia\\
\vspace{1.0cm}

\end{center}

\vspace{2cm}
\begin{center}
\centerline{\bf Abstract} \vspace{.8cm}
\parbox{15.5cm}
{The spin-boson model (or the dissipative two-state system) 
is a  model for the study of dissipation
and decoherence in quantum mechanics.
The   spin-boson model with  Ohmic dissipation  is an integrable theory, 
related to several other integrable systems
including the
anisotropic  Kondo and
resonant level models.
Here we consider  the problem of computing  the
overlaps between  two ground states
corresponding to different 
values of  parameters of the Ohmic  spin-boson Hamiltonian.
We argue that this
can   be understood as a part  of the 
problem of quantizing  the mKdV/sine-Gordon  integrable hierarchy.
The main objective of this work  is to  analyze how the
Anderson orthogonality 
affects   the Yang-Baxter integrable structure underlying the theory. }

\end{center}

\vfill

\end{titlepage}
\newpage

\tableofcontents

\section{Introduction}

\begin{figure}
\centering
\psfrag{A}{${\boldsymbol \mu}_1$}
\psfrag{B}{${\boldsymbol \mu}_2$}
\psfrag{CFT}{CFT}
\psfrag{t}{$y$}
\psfrag{CFT}{CFT}
\psfrag{L}{$L\to\infty$}
\includegraphics[width=6cm]{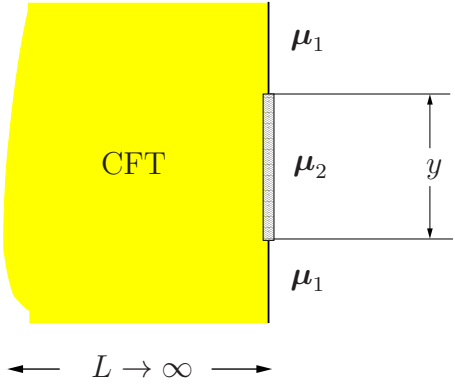}
\caption{An universal part of 
the  overlap modulus  $|\langle\,\Omega_2\,|\,\Omega_1\,\rangle|$
can be extracted from the partition function of a
classical statistical
system with an inhomogeneous boundary. In this picture
 ${\boldsymbol \mu}_1$ and 
${\boldsymbol \mu}_2$ are two numerically different sets of values
of boundary couplings. }
\label{figI3}
\end{figure}

The occurrence  of infrared (IR) divergences is
a central issue for 
quantum field theories and condensed matter systems  which possess   
gapless excitations \cite{Bloch,Anderson}.
Boundary Conformal Fields Theories (CFT) in two-dimensional space-time
provide  an opportunity to gain useful insights into the problem.
In  the simplest set up, with the Euclidean geometry of the half-infinite  plane
$(x,y)$
in which  $x\leq 0$ is treated  as a space coordinate,
the vacuum states 
corresponding to different conformal Boundary Conditions (BC), say ``1'' and ``2'',
are
orthogonal.
More precisely, for a large but finite
system  of space  size $L$  their overlap tends to zero as a power 
$L^{-d_{21}}$, defining an orthogonality exponent $d_{21}$.
This   vacuum overlap can be interpreted as a one-point function 
of the BC  changing
operator ${\cal O}_{21}$ \cite{Cardy},
and therefore 
its   contribution to the  spectral sum   of the  two-point function
$\langle\,{\cal O}^\dagger_{21}(y){\cal O}_{21}(0)\,\rangle$ vanishes, 
as well as   individual   contributions of all overlaps    involving 
conformal descendant states. However, the combined contribution
of the
states organized in conformal towers   turns out to be finite for the
infinitely large system and  gives rise
to the scale-invariant  two-point function 
$\langle\,{\cal O}^\dagger_{21}(y){\cal O}_{12}(0)\,\rangle=|A_{21}|^2\ 
|y|^{-2d_{21}}$. The latter coincides with the ratio
${\cal Z}_{21}(y)/{\cal Z}_{11}$, with     ${\cal Z}_{21}$ standing  for
a  partition function
of the half-infinite  system with    BC ``1''
everywhere except on a part of the boundary of length $y$, where
BC  ``2'' is imposed, whereas the denominator  ${\cal Z}_{11}$ is  a partition function
of the system with  BC ``1''  is imposed  along the whole boundary.
There is generally a normalization ambiguity of the 
numerical coefficient $A_{21}$, and  usually it can be set as 1.
The situation  becomes  more interesting for the so-called  boundary flows, i.e.,
for  a class of two-dimensional  quantum field theories
in which conformal
invariance is broken only by BC depending upon a set of
couplings
${\boldsymbol \mu}=(\mu_1,\mu_2\ldots)$.
Contrary to the conformally invariant  case,
the partition function 
of the system schematically visualized  in  Fig.\ref{figI3}
is a  complicated function of  Euclidean time $y$.
However,
its   $y\to+\infty$ asymptotic  is expected to have the 
form
\bea\label{assa}
\frac{{\cal Z}_{21}(y)}{{\cal Z}_{11}}
=|A_{21}|^2\ y^{-2 d_{21}}\ \re^{-y  \Delta E_{21}} \big(1+o(1)\,\big)\ ,
\eea
where $\Delta E_{21}
\equiv E_0({\boldsymbol \mu}_2)-E_0({\boldsymbol \mu}_1)$
is  the  difference in the  ground state  energies for
the different  sets of 
boundary couplings.
The orthogonality exponent $d_{21}$ and  the prefactor  $A_{21}$ are   
functions of  ${\boldsymbol \mu}_1$ and
${\boldsymbol \mu}_2$ which  virtually define
an universal (i.e. independent on  details of both
IR and ultraviolet (UV) regularizations) part of the vacuum   overlap
\bea\label{osasauias}
\langle\,\Omega_2\,|\, 
\Omega_1\,\rangle \propto   A_{21}\ L^{-d_{21}}\ \ \ \ \ \ {\rm as}\ \ \ L\to\infty\ .
\eea
In common nomenclature,
the modulus of this overlap
is referred as to
the (ground state)  fidelity. 
Throughout this paper, with some abuse of conventional terminology, 
this  term will be used to denote the  scaling function $A_{21}$ 
as well as its  generalization. 
The generalization  deals with the vacuum-vacuum matrix elements
of a bare (unrenormalized) 
boundary field ${\cal O}(y)$
characterized by a  certain 
scaling exponent $D({\cal O})$, 
so that \eqref{osasauias} is  substituted by
\bea\label{osasaus}
\langle\,\Omega_2\,|\,{\cal O}(0)\,|\, \Omega_1\,\rangle \propto  
 A_{21}\big({\cal O}\big)\ 
\varepsilon^{D({\cal O})}\ L^{-d_{21}({\cal O})}\ ,
\eea
defining  both
the IR exponent
$d_{21}({\cal O})$
and  the fidelity $A_{21}\big({\cal O}\big)$
(here $\varepsilon\to 0$ is the lattice spacing, i.e., the UV regulator).

The significance of  the study of  fidelities
is that it may help to better  understand    universal aspects of 
the dynamics after  a local  quantum quench 
in quantum impurity models \cite{Exp1,Exp2,Hubert}.
Such  models are used to mimic the behavior of   small interacting
quantum mechanical systems coupled
to an external environment. In  some cases,
they  display  universality 
which can be described in terms  of
the   boundary flows (see ref.\cite{Afileck} for review of  applications of  the boundary
flows in condensed matter physics).
Here we will  discuss     the so-called 
spin-boson model (or the dissipative two-state system) 
which is a paradigm model for study of dissipation
and decoherence in quantum mechanics \cite{Leggett}.
In the case of 
Ohmic dissipation,
the model   consists of a  single two-state system  coupled
linearly to an infinite bath of harmonic oscillators, and described 
by the Hamiltonian
\bea\label{lsasa}
{\boldsymbol H}=\int_0^\infty{\rm d} k\,  b^\dagger_{k} b_k\ \sigma_0
-J\sigma_1-
h\,\sigma_3
- \sqrt{\frac{g}{2}}\, \int_{0}^\infty \rd k\  (b^\dagger_k+b_k)\  \sigma_3\ ,
\eea
where the Pauli matrixes and $\sigma_0\equiv 1$
describe  the two-state system (``quantum spin''),
$b^{\dagger}_k$ and $b_k$ 
are
phonon creation and annihilation operators such that
$[b_k,b^\dagger _{k'}]= k\, 
\delta(k-k'),\ [b_k,b _{k'}]=[b^\dagger_k,b^\dagger _{k'}]=0$.
The bare tunneling amplitude  between the
eigenstates of $\sigma_1$ is  given by $J$, and
$h$ is an additional bias.
The Ohmic dissipative two-state system is 
related to several other models, 
including 
the anisotropic  Kondo model \cite{AndersonYuval,Chakravarty,Bray}, 
the resonant level model  \cite{Finkel'stein} and 
the inverse square Ising model \cite{ChakravartyRudnick}.

One important issue in  the  spin-boson model is   the
phonon-induced delocalized-localized transition.
Such
a delocalized transition at zero temperature is now
considered as some kind of quantum phase transition.
For  the Hamiltonian \eqref{lsasa}
the quantum transition of
Kosterlitz-Thouless  type occurs at $g=1$. The delocalized region $0<g<1$ 
corresponds to the antiferromagnetic Kondo model, while
the localized region corresponds to the ferromagnetic case.
Here we will consider only the case  $0<g<1$.

As it was pointed out in ref.\cite{Callan}
the Ohmic   bath of harmonic oscillators 
can be interpreted as a
simple bulk CFT  -- the massless Gaussian model.
This allows one to reformulate the spin-boson model in the delocalized regime
as a  boundary flow   problem,
where the parameters of the Hamiltonian $J$ and $h$
play the  r${\hat {\rm o}}$le of the dimensionful  boundary  couplings.
The flow starts from the  Gaussian CFT  with the Neumann (free)  BC and
with the decoupled spin degrees of freedom. For $h=0$, 
the Gaussian field   still satisfies
the Neumann BC in the IR fixed point;
however, the spin 
proves to be completely screened (for details see, e.g., ref.\cite{Afileck}). 

In this work 
we will study the vacuum overlaps      \eqref{osasauias}
where  the vacuums corresponds to  
different sets of the couplings $(J_1,h_1)$ and $(J_2,h_2)$.
The arguments similar to that for the X-ray edge problem 
\cite{Nozieres,Schotte,AffleckLudwig}
leads to the simple  formula for the IR  singularity  exponent
\bea\label{expo}
d_{21}=\frac{g}{4}\ (m_2-m_1)^2\ ,
\eea
where  $m_i=\langle 
\Omega_i\,|\sigma_3\, |\,\Omega_i\,\rangle/\langle \Omega_i\,|\,\Omega_i\,\rangle$.
Despite the lack of a rigorous proof,
there are  strong  indications,
including numerical  results from ref.\cite{Delft1},
that this is an exact relation  for the spin-boson model with $0<g<1$.
The aim of this work
is to  make  steps towards  the exact calculation
of  fidelities.

The paper is organized as follows.
In Sec.\,\ref{Sec2}   
we give a brief account of the basic concepts and facts
and set  up  notations
that will be used in the main body of the text.
Sec.\,\ref{Sec3}
reviews several well-known   techniques for study
of  the orthogonality exponent and fidelities in the spin-boson model.
The purpose of the next two sections is 
to develop a non-perturbative approach for
a calculation  of the
fidelities.
In the absence of  IR divergences 
the Gell-Mann and Low theorem \cite{Gell-Mann}
allows one to express the vacuum overlaps in terms of
the half-infinite time evolution operators
in the interaction picture. 
However, the procedure 
which is based on the adiabatic switch of interaction
generally fails for a system with gapless excitations.
Our approach    is based on an axiomatic determination
of the  fidelities similar 
in philosophy to the form-factor  bootstrap \cite{Fedya}.
In Sec.\,\ref{Sec4} we argue that,
in the case of spin-boson model,
 matrix components of the
half-infinite time evolution operators  can be interpreted as
the quantum Jost operators --
the quantum  counterpart of the
Jost functions for  the pair of
Sturm-Liouville equations.
With this observation, the
calculation  of the
fidelities can   be considered as a part of the
problem of quantizing  the mKdV/sine-Gordon  integrable  hierarchy.
The keystone element of  quantum integrability
 is the Yang-Baxter
type algebras with  commutation relations
defined by  certain  quantum $R$-matrix.
In Sec.\ref{Sec5}, basing on the
results of the
works \cite{Sergei1,BLZ},
we propose a  set
of algebraic relations 
for the quantum Jost operators, which is then translated into a set of functional 
equations imposed on the fidelities.
Currently, the solution of the     system of functional equations  is known  
for the case $h_1=h_2=0$ only.  It was
reported in ref.\cite{Vasseur}.
The last section  of the paper
contains  a few remarks concerning  the fidelities ${\cal A}_{12}(\sigma_s^{(a)})$
corresponding to a family of  bare   operators
\bea\label{aopoas}
\big[\sigma_s^{(a)}(0)\big]_{\rm bare}=
\exp\bigg(a\sqrt{2g}\ 
\int_{0}^\infty\frac{\rd k}{k}\ (b^\dagger_k-b_k)\,\bigg)\ \sigma_s\ ,
\eea
where $\sigma_s\in \big\{\,1,\sigma_3,\frac{1}{2}\, 
(\sigma_1\pm \ri \sigma_2)\,\big\}$ and $a$ is  a real
parameter.
Here
we also present 
formulas for  ${\cal A}_{12}(\sigma_s^{(a)})$ in the case
$h_1=h_2=0$, which generalize  the result of \cite{Vasseur}.
A derivation of 
these formulas  is somewhat technical  and  it remained  beyond the scope of this work.


.

\section{\label{Sec2}\label{prel}Preliminaries}

\subsection{Basics  of  Gaussian  model  with Neumann BC}

We first consider  the Gaussian model
on the half-line  whose dynamics is governed by  
the Hamiltonian,
\bea\label{lkklsla}
{ H}_{\rm free }=\frac{1}{ 4\pi g}\ \int_{-\infty}^0\rd x\
\big(\, \Pi^2+(\partial_x\Phi)^2\, \big)\ ,
\eea
the  Neumann BC, $\partial_x\Phi(x,t)|_{x=0}=0$, and
the canonical commutation relations $[\,\Phi(x),\,\Pi(x')\,]$
$=$
$2\pi \ri\, g \ \delta(x-x'),$ {\it etc}.
The  space of  states      splits up  into  the   Fock spaces ${\cal F}_p$ -- irreps
of the algebra of creation-annihilation operators
\bea\label{uyt}
b_k=\sqrt{\frac{2}{g}}\,\int_{-\infty}^0\frac{{\rm d} x}{4\pi}\ 
\big( (\Pi+\partial_x\Phi)\,\re^{\ri k x}+
(\Pi-\partial_x\Phi)\,\re^{-\ri k x}\big)\ :\ \ \ 
\ \ [b_k,b_{k'}]= k\ \delta(k+k')\, ,
\eea
whose  highest weight vectors   are   
 defined by   the  conditions $b_k\,| \,p\,\rangle=0\ (k>0)$  and
$b_0\, | \, p\,\rangle= \sqrt{2 g}\,  p\, | \,p\,\rangle$.
Throughout  this paper,
we will refer to  $|\,p\,\rangle$ as 
{\it p-vacuums}. 
The  Fock spaces 
are naturally equipped with the inner product defined by the conjugation  
$b_k^\dagger=b_{-k}$ and
$\langle \,p'\,|\,p\,\rangle=\delta_{p',p}$.

\begin{figure}
\centering
\psfrag{y}{$y$}
\psfrag{t}{$t$}
\psfrag{a}{$y_0$}
\psfrag{o}{$0$}
\psfrag{b}{$t_0$}
\includegraphics[width=5cm]{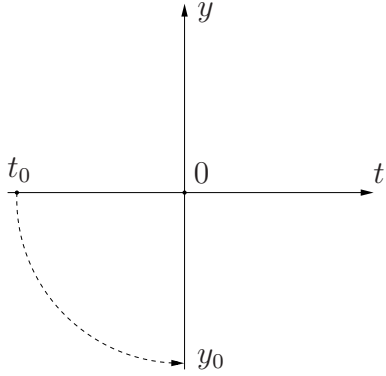}
\caption{The Wick rotation. In the
Euclidean picture,
the model  can be interpreted  as a two-dimensional  classical statistical
system on the    half-plane $\Re  e(z)\leq 0$ with  $z=x+\ri y$.}
\label{fig1}
\end{figure}

Since ${ H}_{\rm free }= \int_0^\infty{\rm d} k\ b_{-k} b_k$, the Hamiltonian
acts invariantly on  each Fock space  $ {\cal F}_p$. Furthermore, all the 
$p$-vacuums 
correspond to the same zero-point  energy, so that
the ground  state of the Gaussian theory with Neumann BC
is a  certain linear combination of the $p$-vacuums.
With the aim   to define the ground state unambiguously, 
it is useful to  consider 
the problem in the Euclidean picture where $t$ is replaced by the
Euclidean time
via the  Wick rotation $t\mapsto y=\re^{\frac{\ri\pi}{2}}\, t$ (see Fig.\,\ref{fig1}).
In the Euclidean picture
all the fields are treated  as functions of $(x,y)$.
Then, the   ground  state
can be defined through the asymptotic
condition
\bea\label{limit}
\lim_{y\to-\infty}\re^{\ri a \Phi(x,y)}\ |\,{\rm vac }\,\rangle=  |\,{\rm vac}\,\rangle\ ,
\eea
which holds true 
for an arbitrary  real parameter $a$.
The exponential fields act between the Fock spaces, $\re^{\ri a\Phi(x,y)} : {\cal F}_p\mapsto {\cal F}_{p+ a}$,
and    we will  always assume the normalization condition
$\langle\,p+a\,|\,\re^{\ri a \Phi(x,y)}\, |\,p\,\rangle=1.$
Thus the ground state  can be written  in the form of a direct integral 
\bea\label{vacu}
|\,{\rm vac}\,\rangle= \int_{-\infty}^\infty\rd p\  |\,p\,\rangle\ .
\eea
The Hilbert space of the model
is given by a linear span 
\bea\label{integ}
{\cal H}=span\big\{\,b_{-k_1}\ldots b_{-k_N}\, |\,{\rm vac} \,\rangle\, \,|\,  
k_i>0\ \&\  N=0,\,1,\ldots\, \big\}\ .
\eea
Note that exponentials  $\re^{\ri a\Phi(x,y)}$ 
act invariantly on ${\cal H}$.

The Gaussian model is manifestly invariant under the transformation
$\Phi(x,y)\mapsto -\Phi(x,y)$, which will be referred below to as 
$C$-conjugation.
 The   corresponding  symmetry operator acts as
\bea\label{CTR}
{\mathbb C}\ :\ \ \ \ \  
{\mathbb C}\, b_k=-b_k\, {\mathbb C}\ ,
\ \ \ \ {\mathbb C}\, |\,{\rm vac}\,\rangle=|\,{\rm vac}\,\rangle\ .
\eea
Another global symmetry  is  the  $T$-invariance. 
The  {\it antiunitary} $T$-transformation acts according to the rule
$\Phi(x,y)\mapsto \Phi(x,-y)$ and the corresponding
symmetry operator ${\mathbb T}$  can be  defined by the relations
\bea\label{CTRA}
{\mathbb T}\ :\ \ \ \ \  {\mathbb T}\, b_k\, {\mathbb T}=-b_{-k}\ ,\ \ \ \ \ 
{\mathbb T}\, |\,{\rm vac}\,\rangle=|\,{\rm vac}\,\rangle\ .
\eea

Finally let us  note that
the  Gaussian  field $\Phi(x,y)$ splits into holomorphic and antiholomorphic components
\bea\label{oassaop}
\Phi(x,y)=\phi(x+\ri y)+\phi(-x+\ri y)\ .
\eea
In fact, 
the  Gaussian  CFT  with Neumann BC
can be    interpreted     as 
a  model   of   a  chiral bose field
on the ``unfolded'' half-infinite line,
whose
Euclidean time evolution, $\phi(x,y)=\phi(x+\ri y)$,
is produced by
the Hamiltonian
\bea
H_{\rm free }
=\frac{1}{2\pi g}\ \int_{-\infty}^{+\infty}\rd x \,\big(\partial_x \phi\big)^2
\eea
through the    commutation relation
\bea\label{com}
[\phi(x_2),\phi(x_1)]=\frac{\ri}{2}\, \pi g\ \sgn(x_2-x_1)\ .
\eea

\subsection{Renormalization in the spin-boson model}

We now turn to  the model of
boundary interaction with  the Hamiltonian
\bea\label{islsasa}
{\boldsymbol H}= { H}_{\rm free }\,\sigma_0
-
({\textstyle\frac{1}{2}}\ { \Pi}_B+h)\, \sigma_3-J\, \sigma_1\ ,
\eea
where $H_{\rm free}$ has been  defined by  eq.\eqref{lkklsla} and 
${ \Pi}_B\equiv { \Pi}(x)|_{x=0}$.  
This  Hamiltonian acts in  the tensor product of the Hilbert space \eqref{integ}
and the   two-dimensional  linear space whose endomorphisms  spanned by
the 
conventional  $2\times 2$
Pauli matrices and $\sigma_0\equiv \begin{pmatrix}1&0\\0&1\end{pmatrix}$. 
The Hamiltonian
is  hermitian for real values 
of the  parameters $J$ and $h$. Without loss of generality
one can  assume that  $J\geq  0$.
In terms of the creation-annihilation operators 
$b_k=b^\dagger_{-k}$ \eqref{uyt} the Hamiltonian 
${\boldsymbol H}$ is given by  eq.\eqref{lsasa}.
In this form it  occurs
as a particular realization of the Caldeira-Leggett Hamiltonian \cite{Leggett}.
The model  is usually referred   to as the spin-boson model or  
dissipative two-level system and used to mimic the behavior of 
dissipative particle confined in a double-well potential. 

The spin-boson model  needs renormalization.
The Hamiltonian \eqref{lsasa} has to be 
equipped with the ultraviolet cut-off $\Lambda$ and consistent removal of the
UV divergences requires the bare coupling 
constants $J$ and $g$  be given a dependence of the cut-off
according to Renormalization Group (RG) flow equations. 
There  exists a  RG scheme where
\bea\label{RG}
\Lambda\ \frac{\rd J}{\rd \Lambda}=
g\ J\ ,\ \ \ \ \ \  \Lambda\ \frac{\rd g}{\rd \Lambda}= 0\ ,
\eea
and because of this
one can   substitute the bare coupling $J$  by the
RG invariant energy scale 
\bea\label{asopsa}
E^\star=const\ \Lambda\ \Big(\frac{J}{\Lambda}\Big)^{\frac{1}{1-g}}\ .
\eea
The latter  is  defined up to a multiplicative
$g$-dependent  constant and  
usually referred to as Kondo temperature in the context of the 
anisotropic Kondo model. 
The parameter $h$ is interpreted as an external  magnetic field applied to the
impurity  spin.
It is often  convenient to specify   the Kondo temperature as
\bea\label{oaspsaos}
{E^{\star}}=-\bigg[\frac{\partial^2}{\partial h^2}E_0(J,h)\bigg]^{-1}_{h=0}\ ,
\eea
where $E_0$ stands for the  ground state energy  
considered as a function  of the bare coupling $J$ and  $h$.

Perhaps the simplest  way to 
understand the  renormalization scheme \eqref{RG}
is based on an  alternative 
form of the Hamiltonian ${\boldsymbol H}$ \eqref{islsasa}.
As it was already mentioned,
the  Gaussian theory with Neumann  BC can be 
interpreted  as a model of a free chiral  boson.
Eq.\eqref{oassaop} implies that  ${ \Pi}_B=-2\, \partial_x\phi(0)$
and, therefore,
the canonical
transformation  ${\boldsymbol  H}
\mapsto {\boldsymbol U}^\dagger {\boldsymbol H} {\boldsymbol  U}$
 with ${\boldsymbol U}=\exp\big(\ri\sigma_3\phi(0)\big)$
brings the Hamiltonian \eqref{islsasa} to the form
\bea\label{aosaps}
{\boldsymbol H }_{\phi}={\boldsymbol U}^\dagger {\boldsymbol H} {\boldsymbol  U}=
\frac{1}{2\pi g}\ \int_{-\infty}^{+\infty}\rd x
 \,\big(\partial_x \phi\big)^2\ \sigma_0-h\, \sigma_3-\mu\,
\big(\re^{+2\ri\phi(0)}\, \sigma_-+\re^{-2\ri\phi(0)}\, \sigma_+\big)
\eea
with $\sigma_\pm=
\frac{1}{2}(\sigma_1\pm \ri \sigma_2)$ and
 $[\phi(x),\phi(x')]=\frac{\ri}{2}\, \pi g\ \sgn(x-x')$.
Notice  that in  eq.\eqref{aosaps}
it is assumed that 
the Hamiltonian is expressed in terms of the {\it renormalized}
exponential operators 
so that the bare coupling  $J=J(\Lambda)$ is substituted by the
RG-invariant    $\mu$.
In order to assign a  precise meaning to  the renormalized coupling   one needs to 
specify  a normalization condition
for the renormalized exponentials.
In fact, we have already accept the condition
$\langle\,p+a\,|\,\re^{2\ri a \phi(y)}\, |\,p\,\rangle=1$. This
sets a  value of
the  leading term of  the  Euclidean   operator product expansion
\bea
\re^{2\ri \phi(y)}\, 
\re^{-2\ri\phi (-y)}\to\ (+1)\times (2y)^{-2 g }\ \ 
\ \ \ \ {\rm as}\ \ y=\ri t\to 0\ \ \ \ \ \
(y>0)\ ,
\eea
where $\re^{-2\ri  \phi(-y)}=\big(\re^{2\ri \phi(y)}\big)^\dagger$.
Thus the renormalized  coupling $\mu$ has 
the dimension of  $[\it energy\,]^{1-g}$,
i.e., $\mu= J\Lambda^{g}$  and $\mu\propto (E^{\star})^{1-g}$,
which yields  formula \eqref{asopsa}.
The exact   $E^\star-\mu$ relation  can be extracted from the results 
of the Bethe ansatz solution of the anisotropic
Kondo model \cite{TsvelickWiegmann,AndreiLowenstein} (see also \cite{BLZ}):
\bea\label{E-mu}
E^\star=(1-g)\ 
\frac{\sqrt{\pi}\Gamma\big(1+\frac{g}{2(1-g)}\big) }{
\Gamma\big(\frac{1}{2}+\frac{g}{2(1-g)}\big)}\  
\big(\,\Gamma(1-g)\,\mu\,\big)^{\frac{1}{1-g}}\ .
\eea

\subsection{\label{Sec2.3}Fidelities $A_{21}({\cal O}_\pm)$ and
  $A_{21}(\sigma_{0,3})$}

\begin{figure}
\centering
\psfrag{A}{$(E^\star_1,h_1)$}
\psfrag{B}{$(E^\star_2,h_2)$}
\psfrag{O}{${\cal O}$}
\psfrag{OD}{${\cal O}^\dagger$}
\psfrag{t}{$y$}
\psfrag{CFT}{Gaussian CFT}
\includegraphics[width=6cm]{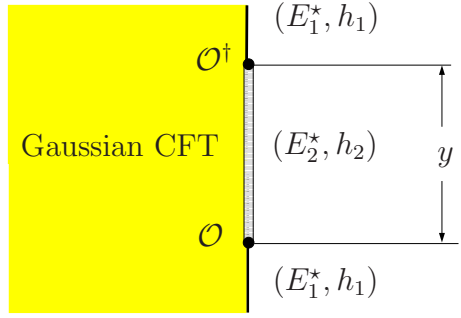}
\caption{
 An universal part of
$|\langle\,\Omega_2\,|\,{\cal O}(0)\, |\,\Omega_1\,\rangle|$
can be extracted from the partition function of the
system with inhomogeneous boundary,  and with  the  
insertion of a  pair
of hermitian conjugate boundary fields
${\cal O}$  and ${\cal O}^\dagger$.
}
\label{figI2}
\end{figure}

Let us slightly generalize  the setting  from
the introduction  and consider 
the  partition function ${\cal Z}_{21}\big({\cal O}|y\big)$
of  the  half-infinite system
with  the  insertion of a  pair
of hermitian conjugate boundary fields
${\cal O}$  and ${\cal O}^\dagger$
at the  ends of the boundary segment 
where  BC 
depends upon  two energy  scales   $(E^\star_2,h_2)$ (see Fig.\,\ref{figI2}). 
In what follows 
we will make use of the notation
\bea\label{aopasopsapas}
{\bar {\cal Z}}_{21}\big({\cal O}|y\big)=
\frac{{\cal Z}_{21}({\cal O}|y)}{{\cal Z}_{11}}\ ,
\eea
where  ${\cal Z}_{11}$ 
is the partition function of the system without any boundary insertions and
whose BC is defined  by
 $(E^\star_1,h_1)$ homogeneously along the whole boundary.
Of course, ${\bar {\cal Z}}_{21}({\cal O}|y)$ 
is a  complicated function  and it is a challenging problem
to compute it in a    compact and manageable form
even for  
the simplest boundary fields.
However, since the physics of 
the model  \eqref{islsasa}  is
well understood now,  one can  make  some 
general predictions regarding its behavior for
small and large values of  $y$.
In what follows we will mostly discuss the simplest case with
${\cal O}$ given by  the diagonal matrixes: $\sigma_0$,  $\sigma_3$, or
\bea
{\cal O}_\pm=\frac{1}{2}\ (\sigma_0\pm \sigma_3)\ .
\eea
Since the model  is asymptotically free at short distances,
the effect of the boundary energy scales $(E^\star_2,h_2)$
from the interval $(0,y)$ becomes negligible wherein 
its size shrinks to zero,
and therefore
\bea
\lim_{y\to 0^+}{\bar {\cal Z}}_{21}\big({\cal O}|y\big)
=\begin{cases}
1\ \ \ \ \ &{\rm for}\ \ \  {\cal O}=\sigma_0,\,\sigma_3\\
{\textstyle \frac{1}{2}}\ \big(1\pm  m_1\big)
\ \ \ \ \ &{\rm for}\ \ \  {\cal O}={\cal O}_\pm
\end{cases}\ .
\eea
Here we  use  
\bea
m=
\frac{\langle\, \Omega\,|\, 
\sigma_3\,|\,\Omega \rangle}{\langle\, \Omega\,| \,\Omega \rangle}
\eea
and its  subscript ``1''  shows that the   expectation value is 
taken over the ground  state $| \,\Omega_1 \rangle$
corresponding to  $(E^\star_1,h_1)$. 
Note that $m$ coincides with the 
impurity magnetization in the context of the anisotropic Kondo model.
It can be expressed in terms of 
the exact ground state energy $E_0$ of the Hamiltonian \eqref{islsasa} as
\bea\label{ospasps}
m=-\bigg(\frac{\partial E_0}{\partial h}\bigg)_{E^\star}\ .
\eea
The spin degrees of freedom are freezing at large distances so that
the large-$y$ asymptotic of
${\bar {\cal Z}}_{21}\big({\cal O}|y\big)$
has a form  
similar to \eqref{assa} with   the same  exponent 
$d_{21}=d_{21}\big( \{E^\star_i,h_i\}\big)$
for any  choice of the diagonal matrix ${\cal O}\in
\big\{\sigma_0,\sigma_3,{\cal O}_\pm\big\}$:
\bea\label{kssus}
{\bar {\cal Z}}_{21}\big({\cal O}|y\big)
=
\big|A_{21}\big({\cal O}\big)\big|^2\ \ 
y^{-2  d_{21}}\
   \re^{-y \Delta E_{21}}\ \big(1+o(1)\big)\ \ 
\  {\rm as}\ \  y\to+\infty\  ,
\eea
where $\Delta E_{21}=E_0(E^\star_2,h_2)-E_0(E^\star_1,h_1)$.
Although  \eqref{kssus}  defines 
$A_{21}\big({\cal O}_{\pm}\big)$ and $A_{21}\big(\sigma_{0,3}\big)$
in absolute value,
their relative phases
are  dictated  by the relations
\bea\label{aoisaisa}
A_{21}\big(\sigma_s\big)=A_{21}\big({\cal O}_{+}\big)+(-1)^s\
A_{21}\big({\cal O}_{-}\big)\ \ \ \ \ \  \ \ \ \ \ (s=0,3)\ .
\eea
In what follows, we will call
$A_{21}\big({\cal O}\big)$
as ``fidelities''
and treat them   as scaling  functions depending on
the    magnetic moments $m_1$, $m_2$ \eqref{ospasps} as well as  the pair of
the Kondo temperatures $E^\star_1$, $E^\star_2$
\eqref{oaspsaos}. In fact, since
they are  dimensionful quantities, they
depend  non-trivially   upon
the   dimensionless   variables   $m_1$, $m_2$ and $\alpha\equiv
\log(E^\star_2/E^\star_1)$ only.

\section{\label{Sec3}Mean field, perturbation theory and 
Toulouse limit}

In this     section we outline several common approaches for
study
of the fidelities in the spin-boson model.

\subsection{\label{Sec3.1}Mean field approximation}

We start with the model which is considerably simpler than  the spin-boson model.
The simplified  
Hamiltonian is obtained from ${\boldsymbol H}$ \eqref{islsasa} through the  substitution
of the Pauli matrix  $\sigma_3$ by a constant $m$.
It splits into two non-interacting parts: 
\bea\label{isasa}
 H_1= { H}_{\rm free }-\frac{m}{2}\  { \Pi}_B\ \ \ :\ \  \ {\cal H}\mapsto {\cal H}
\eea 
and a  $2\times 2$ matrix  
${\boldsymbol  H}_2= -h\,\sigma_3-J\, \sigma_1$.
This can be thought  of  as a  mean field 
approximation,  with   the value of  $m=m(h)$  
given by the relation   $m=-\frac {\partial E_0}{\partial h}$,
where $E_0=-\sqrt{J^2+h^2}$ is the lowest eigenvalue of ${\boldsymbol  H}_2$.

To construct the vacuum state  for  the mean field Hamiltonian
one can  use the  interaction picture  with the  term  $\propto m$ in \eqref{isasa} 
is treated as an interaction. 
The unitary operator $U(t,t_0)=\re^{\ri  H_{\rm free } (t-t_0)}\ \re^{\ri H_1(t-t_0) }$ 
can be calculated explicitly in this case:
\bea
U(t,t_0)=\re^{\frac{\ri m}{2}\Phi_B(t)}\ \re^{-\frac{\ri m}{2}\Phi_B(t_0)}\ ,
\eea
where ${ \Phi}_B(t)=\re^{\ri 
{ H}_{\rm free} t}\,  { \Phi(0,0)}\,
\, \re^{-\ri { H}_{\rm free} t}$.
The 
vacuum   for \eqref{isasa}
is  obtained 
through  the  
Euclidean time evolution of the state 
$|\,{\rm vac}\,\rangle$ \eqref{vacu}.
Thus the vacuum state for the whole mean field  Hamiltonian is given by
\bea\label{ksasaiasi}
|\, \Omega\,\rangle=
 \re^{\frac{\ri m}{2}\, \Phi_B(0)}\
\lim_{t_0\to \infty}\,
\re^{-\frac{\ri m}{2}\, \Phi_B(t_0)}\  |\, {\rm vac }\, \rangle
\otimes
\frac{1}{\sqrt{2}}\,
\begin{pmatrix}
\sqrt{1+m}\\
\sqrt{1-m}
\end{pmatrix}\ ,
\eea
where the limit is taken along imaginary time direction 
$y_0=\ri\, t_0\to -\infty$ as it is shown  in Fig.\,\ref{fig1}.  
Taking into account   the defining property \eqref{limit}  of the state $|\,{\rm vac}\,\rangle$,
one obtains 
\bea\label{ksiasi}
|\, \Omega\,\rangle =T_+\ |\, {\rm vac }\, \rangle\otimes|\,\uparrow\,\rangle+
T_-\ |\, {\rm vac }\, \rangle\otimes|\,\downarrow\,\rangle\ ,
\eea
where
\bea\label{osapas}
T_\pm=
\sqrt{\frac{1\pm m}{2}}\  \re^{\frac{\ri m}{2}\, \Phi_B(0)}\ ,
\eea
and we use the common notation for $\sigma_3$-eigenvectors.
Similarly one has
\bea\label{ksiasib}
\langle\, \Omega\,|=\langle\, {\rm vac}\, |\, Q_+\, \otimes\langle\,\uparrow\,|+
\langle\, {\rm vac}\, |\, Q_-\, \otimes\langle\,\downarrow\,|\nonumber
\eea
with
\bea\label{osapasa}
Q_\pm=\big(T_\pm\big)^\dagger=
\sqrt{\frac{1\pm m}{2}}\  \re^{-\frac{\ri m}{2}\, \Phi_B(0)}\ .
\eea

It is easy to see now that the mean field approximation yields the relation
\bea
{\cal Z}_{21}({\cal O}_\pm|y)\propto 
\langle\, {\rm vac }\,|\, \re^{- \ri \omega\Phi_B(y)}\, 
\re^{+\ri \omega\Phi_B(0)}\,|\,{\rm vac}\,\rangle\ \ \ {\rm with}\ \ \ \omega=
{\textstyle\frac{1}{2}}\,(m_1-m_2)
\eea
and therefore 
to eq.\eqref{expo} for the IR singularity exponent.
As for the fidelities  $A_{21}\big({\cal O}_\pm\big)$,
it is  worth to note the relation 
\bea\label{oaisau}
\langle\, p'\,|\,Q^{(2)}_{\pm}\,  T^{(1)}_{\pm}\,|\, p\,\rangle=
A_{21}\big({\cal O}_\pm\big)
\ \delta_{2p'-2p,m_1-m_2 } \  .
\eea
Within the mean field approximation 
$A_{21}\big({\cal O}_\pm\big)=
\frac{1}{2}\sqrt{(1\pm m_2)(1\pm m_1)}$,
which  is found to be an adequate approximation  as $g\to 0$.
Contrary to the fidelities, the formula  $d_{21}=\frac{g}{4}(m_2-m_1)^2$
is expected to be  exact  as $m_1$ and $m_2$ are understood as    vacuum
expectation values of $\sigma_3$ in the
spin-boson model.
Evidences  in  its favor are presented in the next  two subsections.

\subsection{\label{RPT}Renormalized perturbation theory}

Similarly to the case $h_1=h_2=0$ considered in ref.\cite{Vasseur},
the fidelities  $A_{21}\big({\cal O}_{\pm}\big)$ 
with $h_1,h_2\not=0$ can be calculated by
means of the renormalized perturbation theory in coupling $g$ 
for the Hamiltonian \eqref{lsasa}.
The result of  perturbative calculations
turn out  to be in   agreement with  the orthogonality
exponent $d_{21}=\frac{g}{4}(m_2-m_1)^2$.
This ensure that 
$A_{21}\big({\cal O}_{\pm}\big)$ can be written in the form
\bea\label{un}
A_{21}\big({\cal O}_{\pm}\big)
=
\big(E^\star_2\big)^{\frac{g}{4}(m_2\mp 1)( m_1-m_2)}\
\big(E^\star_1\big)^{\frac{g}{4}(m_1\mp 1)(m_2-m_1)}\ \ 
{\cal  A}_{\pm }(m_2,m_1|\,\alpha)\ ,
\eea
where 
\bea\label{gamma}
\alpha\equiv  \log\big(E^\star_2/E^\star_1\big)\ ,
\eea
and
the prefactor has a dimension 
of $[{\it energy}\,]^{-d_{21}}$, so that
 ${\cal  A}_{\pm }$ are    dimensionless amplitudes.
Notice  that in the case $\mu_2=\mu_1$, $h_2=h_1$, 
the orthogonality exponent vanishes and ${\cal  A}_{\pm }$
should satisfy an  exact relation
\bea\label{normab}
{\cal A}_{\pm}(m,m\,|\,0\,)=\frac{1}{2}\, (1\pm  m)\ .
\eea

It can be shown with somewhat cumbersome but straightforward effort that
\bea\label{aspass}
&&{\cal A}_{+}(m_2,m_1\,|\,\alpha)=
\frac{1}{2}\ \sqrt{(1+m_2)(1+m_1)}\ \  \big(2\re^{\gamma_E}
\big)^{-\frac{g}{4}(m_1-m_2)^2}\\
&&\ \ \ \  \times\ \,
\big(1-m_2^2\big)^{\frac{g}{8}(1-m_2)(m_1-m_2)}\
\big(1-m_1^2\big)^{\frac{g}{8}(1-m_1)(m_2-m_1)}
\nonumber
\\
&&\ \ \ \  \times\ \,
\bigg(1+
\frac{g}{4}\ \Big(\,(1-m_1)^2+(1-m_2)^2- (1-m_1)(1-m_2)\,
{\delta}\, \coth\frac{{\delta}}{2}\, \Big)
+O(g^2)\,\bigg)\nonumber
\eea
and
\bea\label{relsus}
{\cal A}_{-}(m_2,m_1\,|\,\alpha)=   
{\cal A}_{+}(-m_2,-m_1\,|\,\alpha)\ . 
\eea
Here $\gamma_E$ stands for the Euler constant, and
 we use
$\delta\equiv\alpha+\frac{1}{2}\log\big(\frac{1-m_1^2}{1-m_2^2}\big)$.
The quoted result shows
that the perturbative   amplitudes ${\cal A}_{\pm}$
are multivalued functions of the complex 
variables $(m_2,m_1|\,\alpha)$. However  their
overall phases
can be
chosen in such a way that they
are {  real analytic} within the domain
\bea\label{phys}
{\mathbb  D}\equiv
\big\{\, (m_1,m_2\,|\,\alpha)\ :\ m_{1,2}\in (-1,1)\,;\  \alpha\in {\mathbb R}\,
\big\}\ .
\eea
It is also easy to see that  ${\cal A}_{\pm}$
satisfy the condition
\bea\label{fornf}
{\cal A}_{\pm}(m_2,m_1\,|\,\alpha)=
{\cal A}_{\pm}(m_1,m_2\,|-\alpha)\ .
\eea

Besides, the use of 
the perturbation theory allows one to determine
the relation between
$m$ and dimensionless  ratio $h/E^\star$ in a form of a power series in $g$.
In particular, to  the first order, one has
\bea\label{alasalsi}
\frac{h}{E^\star}=\frac{m}{\sqrt{1-m^2}}\ \Big(1-\frac{g}{2}\ 
\log(1-m^2)+O(g^2)\,\Big)\ .
\eea
With this formula the perturbative amplitudes
can be    expressed in terms of $h_i/E^\star_i\ (i=1,2)$.
Notice that, from the Bethe ansatz solution of the
anisotropic Kondo  model, it is known that 
\cite{TsvelickWiegmann,AndreiLowenstein}
\bea\label{ioaiso}
m=\frac{\ri}{ 2\pi}\
\int_{-\infty}^{\infty}\rd\omega\, \lambda^{\frac{\ri\omega}{1-g}}
\ \  \frac{M(\omega)}{ \omega+\ri 0}
\ ,\   \ \ M(\omega)=\frac{\Gamma(1-\frac{\ri\omega}{ 2(1-g)})
\Gamma(\frac{1}{ 2}+\frac{\ri\omega}{ 2})}{
 \sqrt{\pi}\ 
\Gamma(1-\frac{\ri\omega g}{ 2(1-g)})}\, 
\Big( \Gamma(1-g)\Big)^{\frac{\ri \omega}{1-g}}\ ,
\eea
where 
\bea
\lambda=
\frac{1}{\Gamma(1-g)}\ \Bigg[\frac{\Gamma\big(\frac{1}{2}+\frac{g}{2(1-g)}\big)}
{\sqrt{\pi}(1-g)\Gamma\big(1+\frac{g}{2(1-g)}\big) }\Bigg]^{1-g}\ 
\Big(\frac{E^\star}{h}\Big)^{1-g}\ .
\eea
This remarkable  exact result   implies the following general 
structure of the
perturbative expansion \eqref{alasalsi}:
\bea
\frac{h}{E^\star}=m\,(1-m^2)^{-\frac{1}{2-2g}}\,\bigg(1+\sum_{n=2}^\infty g^n\sum_{l=1}^{n-1}
c_{n l}\,  m^{2l}\,\bigg)\ ,
\eea
where $c_{nl}$ are some numerical coefficients.

\subsection{\label{Sec3.3}Toulouse limit}

In the  case $g=\frac{1}{2}$, which is sometimes  referred as to 
``Toulouse limit'',
the
Hamiltonian ${\boldsymbol H}_\phi$  \eqref{aosaps}
can be fermionized    in  terms of  the chiral complex fermion field
\bea\label{appasos}
{\boldsymbol H}_{\rm Toul}=
\frac{1}{2\pi\ri }\int_{-\infty}^\infty\rd x\ \psi^\dagger\partial_x\psi
-\mu\, \big(\, {\hat d}^\dagger  \psi(0) +
\psi^\dagger(0) {\hat d}\, \big)+
h\, ({\hat d}^\dagger {\hat d}- {\hat d}{\hat  d}^\dagger )\ ,
\eea
where $\{\psi^\dagger(x), \psi(x')\}=\delta(x-x'),\ 
\{{\hat d}^\dagger, {\hat d}\}=1$, {\it etc}.
Equivalently, the model can be understood as a
boundary flow for  the 
Dirac fermion, massless in the bulk.
In order to construct an Euclidean action  governing 
this boundary flow,
we define ${\bar \psi}(x,y)\equiv{ \psi}(-x,y)$, so that the Hamiltonian
\eqref{appasos} for $\mu=h=0$  takes the form
$\frac{1}{2\pi\ri }\int_{-\infty}^0\rd x\
\big(\psi^\dagger\partial_x\psi-
{\bar \psi}^\dagger\partial_x{\bar\psi}\big)
$.  The  fields ${ \psi}$ and ${\bar  \psi}$ are interpreted now
as   components of  the  massless
Dirac fermion, both   defined   on
the  half-infinite lane $x\leq 0$ and satisfying  
the  bulk equations of motion
$\partial_{\bar z}\psi=\partial_z{\bar \psi}=0$
with  $z=x+\ri y$, ${\bar z}=x-\ri y$.
The complex  fermions can be substituted  by two Majorana-Weyl fermions:
$\psi=\frac{\psi_1+\ri \psi_2}{\sqrt{2}}$,
${\bar\psi}=\frac{{\bar \psi}_1+\ri {\bar \psi}_2}{\sqrt{2}}$.
Each of the real fermions $(\psi_j,{\bar \psi}_j)$ satisfies the free BC,
$\big(\psi_j-{\bar\psi}_j\big)|_{x=0}=0$ and, as it was explained
in ref.\cite{CZ},
should  be described
by means  of the action $
{\cal A}_{\rm MW}[\psi_j,{\bar \psi}_j, a_j]$
which involves additional boundary fermionic degree of freedom $a_j=a_j(y)$:
\bea
{\cal A}_{\rm MW} = \frac{1}{2\pi{\rm i}}\int_{-\infty}^\infty {\rm d}y
 \int_{-\infty}^0 {\rm d}x\, \left(\psi_j{\partial}_{\bar z}\psi_j-
\bar{\psi}_j\partial_z\bar{\psi}_j\right)+\frac{1}{2}
 \int_{-\infty}^\infty {\rm d}y
 \Big( \frac{1}{2\pi{\rm i} } (\psi_j\bar{\psi_j})|_{x=0}+a_j\partial_y{a_j}
\Big).
\eea
An  Euclidean  action, corresponding to the Hamiltonian \eqref{appasos}
with non-vanishing couplings $\mu$ and $h$,
is given by \cite{CZ}\footnote{
Although in ref.\cite{CZ} 
the case of a single Majorana-Weyl fermion
was only
discussed,
Eq.\eqref{hassya} is   an apparent consequence 
of
the Chatterjee-Zamolodchikov result.}
\bea\label{hassya}
{\cal A}_{\rm Toul}=\sum_{j=1}^2{\cal A}_{\rm MW}[\psi_j,{\bar\psi_j},a_j]+
\int_{-\infty}^\infty \rd y\ \Big(\,\mu\,  {\mathfrak S}_B(y)-
h\  {\mathfrak M}_B(y)\,\Big)\  ,
\eea
where
\bea\label{aisosai}
{\mathfrak S}_B=\frac{\ri}{2}\ \sum_{j=1}^2a_j\,
\big(\psi_j+{\bar \psi}_j\big)|_{x=0}\ ,\ \ \ \  \ \
{\mathfrak M}_B=2\ri\ a_1a_2\ .
\eea
Clearly,   the      Grassmannian boundary fields
$d=\frac{a_1+\ri a_2}{\sqrt{2}}$,
$d^*=\frac{a_1-\ri a_2}{\sqrt{2}}$
in the path integral formalism correspond to the
nilpotent  operators ${\hat d}$ and ${\hat  d}^\dagger$ 
in the Hamiltonian picture.

The general solution of the bulk equations,  
$\partial_{\bar z}\psi=\partial_z{\bar \psi}=0$, are given by 
the  Fourier integrals
\bea\label{apopsa}
\psi=\int_{-\infty}^\infty\rd k\ c_k\ \re^{\ri  k z}\ ,\ \ \ \
{\bar \psi}=\int_{-\infty}^\infty\rd k\ {\bar c}_k\ \re^{-\ri k{\bar z} }\ .
\eea
In their turn, the boundary equations of motion
corresponding to the  action \eqref{hassya}
allows one to express  the Fourier modes  ${\bar c}_k$ and 
the Heisenberg operator ${\hat d}(y)$  in terms  of $c_k$:
\bea\label{aiosia}
{\bar c}_k=\re^{2\ri\delta(k)}\ c_k\ \ \ \ \ \ \ {\rm with}\ \ \ \ \ 
\re^{2\ri\delta(k)}=
\frac{k-\ri \pi\mu^2-2h}{k+\ri\pi\mu^2-2h}
\eea
and 
\bea\label{aopsopas}
{\hat d}(y)=\ri\  \mu\
\int_{-\infty}^\infty\rd k\ \frac{c_k\ \re^{-ky}}{k+\ri\pi\mu^2-2h}\ .
\eea
With this, it is straightforward to compute
the Euclidean propagator
 for  the complex boundary fermions:
\bea
\langle\, d^{* }(y)\, d(0)\,\rangle=D(y|-h)\ \Theta(y)-\Theta(-y)\ D(y|h)\ ,
\eea
where $\Theta(y)=\frac{1}{2}(1+{\rm sign}(y)\big)$ and
\bea\label{oaspasaso}
D(y|h)=\mu^2 \int_0^\infty\rd k\ \frac{\re^{-k|y|}}{(k- 2 h)^2+
\pi^2\mu^4}\to 
\begin{cases}
\frac{1+ m}{2}\ \ \ &{\rm as}\ \ \ y\to 0\\
\frac{\cos^2(\frac{\pi m}{2})}{4 E^\star y}\ \ \ &{\rm as}\ \ \ y\to \infty
\end{cases}\ .
\eea
In the last formula we
use the notations
 $E^\star=(\frac{\pi\mu}{2})^2$ and
\bea\label{ospoaspspo}
m=\frac{2}{\pi}\ {\rm arctan}
\Big(\frac{\pi h}{2 E^\star}\Big)\ ,
\eea
which are  consistent  with general  relations \eqref{E-mu} and \eqref{ioaiso}
taken at $g=\frac{1}{2}$.

In order to calculate the 
orthogonality exponent,
one needs to  introduce  an explicit IR regularization. 
Let us restrict  values of the space coordinate $x$ to the segment $[-L,0]$ 
and choose  the 
free BC at $x=-L$: 
$\big(\psi-{\bar\psi}\big)|_{x=-L}=\big( \psi^\dagger-
{\bar\psi}^\dagger\big)|_{x=-L}$, or equivalently,
${\tilde c}_k\ \re^{\ri L k}= c_k\ \re^{-\ri L k}.$
Taking this together with
eq.\eqref{aiosia}, one obtains the quantization condition
$2Lk_n+2\,\delta(k_n)=2\pi n\  (n\in{\mathbb Z}),$ so that
Fourier integral expansions \eqref{apopsa} should be  substituted by  discrete sums 
$\psi=\sum_{n=-\infty}^\infty {\tt  c}_n\ \re^{\ri k_n(x-t)}$, and similar for 
${\bar \psi}$.
Note that the real-time evolution of the fermion modes ${\tt  c}_n$ 
is produced by the Hamiltonian
$H=\sum_{n=-\infty}^{+\infty}k_n\,{\tt c}_n^\dagger {\tt c}_n$ through
the canonical commutation relations 
$\{{\tt c}_n^\dagger,{\tt c}_{n'}\}=\delta_{n,n'},\  
{\tt c}^2_n=({\tt c}^\dagger_n)^2=0$.
A ground state of the system of fermionic  oscillators
is defined by the requirement
that all energy levels bellow the Fermi level $k_F=0$ are
occupied.
In this situation, according to Anderson \cite{Anderson},
the orthogonality exponent for the ground state  overlap 
$\langle\,\Omega_2\,|\,\Omega_1\,\rangle$
is determined by the
the difference of the phase shifts
at the Fermi level:
\bea\label{jaisuas}
d_{21}\big|_{g=\frac{1}{2}}
=\frac{1}{2\pi^2}\ \big(\delta_2(0)-\delta_1(0)\big)^2\ .
\eea
In its turn, 
the phase shift $\delta(k)$ \eqref{aiosia} 
at $k=0$ can be written  in terms
of $m$ \eqref{ospoaspspo} as
$\delta(0)=\frac{\pi}{2}\ (1-m)$,
and therefore
$d_{21}=
\frac{1}{8}\ (m_2-m_1)^2$.
This coincides with \eqref{expo} specialized at   $g=\frac{1}{2}$.

It is useful to note that
the orthogonality exponent can be written in the form
\bea\label{aoipsaops}
d_{21}=
\frac{1}{2}\ (q_1-q_2)^2\ ,
\eea
where $q_i=\langle\Omega_i\, |\, {\hat q}\, |\,\Omega_i\,\rangle$ 
are   vacuum expectation values of the operator
${\hat q}=\sum_{n=-\infty}^{+\infty}c_n^\dagger c_n$.
Numerical results
presented in  ref.\cite{Delft1} suggest that
eq.\eqref{aoipsaops} remains valid for $g\not=\frac{1}{2}$.
For their calculations, the authors used
the   resonant level model  \cite{Finkel'stein}
whose Hamiltonian is obtained by adding a   four-fermion interaction
term to  ${\boldsymbol H}_{\rm Toul}$ \eqref{appasos},
\bea\label{appasososi}
{\boldsymbol H}_{\rm RLM}=
{\boldsymbol H}_{\rm Toul}+u\,({\hat d}^\dagger {\hat d}-
{\hat d}^{\dagger}{\hat d})\ :\psi^\dagger\psi(0):\ .
\eea
As it is well known (see, e.g., \cite{Gogolin,BoulatSaleur}), this is
 a fermionic version of the Hamiltonian  
${\boldsymbol H}_{\phi}$ for general values of  $g\in [0,1]$.
The difference $q_1-q_2$  in \eqref{aoipsaops} 
was referred  in ref.\cite{Delft1} as to
the ``displaced charge'', which   is, in a sense,
the difference between vacuum expectation values of the
charges associated with the global $U(1)$ symmetry of the resonant level model.
We may  now  note that  formulae  \eqref{expo} and \eqref{aoipsaops} for the
orthogonality exponent coincides provided
\bea
q_1-q_2=\sqrt{\frac{g}{2}}\ (m_2-m_1)\ .
\eea
The last  relation indeed holds true for the   resonant level model  
(see  Chapter 28IV.2 in ref.\cite{Gogolin} for details).

Returning to the spin-boson model at the Toulouse limit,
we acknowledge the relations between 
the  normalized partition functions
${\bar {\cal Z}}_{11}(\sigma_{3,1}|y)\equiv
{\bar {\cal Z}}_{21}(\sigma_{3,1}|y)|_{E^\star_2=E^\star_1, h_2=h_1}$ and
two-point functions of the boundary fields \eqref{aisosai}:
\bea\label{psoapaspas}
{\bar {\cal Z}}_{11}\big(\sigma_3|y\big)\big|_{g=\frac{1}{2}}&=&
\langle\,\mathfrak{M}_B(y)\,\mathfrak{M}_B(0)\,\rangle\\
{\bar {\cal Z}}_{11}\big(\sigma_1|y\big)\big|_{g=\frac{1}{2}}&=&
\langle\,\mathfrak{S}_B(y)\,\mathfrak{S}_B(0)\,\rangle\ .\nonumber
\eea
The boundary equations of motion  corresponding to  the action
\eqref{hassya}, allows  one to represent the  Heisenberg   operators
${\hat {\mathfrak{M}}}_B(y)$ and ${\hat {\mathfrak{S}}}_B(y)$ 
in the form of    normal-ordered
combinations of  ${\hat d}(y)$  and its Hermitian conjugates: 
\bea\label{aopsaoas}
{\hat {\mathfrak{M}}}_B&=&2:{\hat d}{\hat d}^\dagger :+m\\
{\hat {\mathfrak{S}}}_B&=&\mu^{-1}\ 
:\big(\, ({\partial_y {\hat d}})\,{\hat d}^\dagger+
(\partial_y{\hat {d}}^\dagger) 
{\hat d}-4h\, {\hat d}^\dagger 
{\hat d}\big):+\langle\, \mathfrak{S}_B\,\rangle\ ,\nonumber
\eea
where the vacuum expectation value,  
\bea
\langle\,\mathfrak{S}_B\,\rangle={\textstyle \frac{2}{\pi}}\ \sqrt{E^\star}\ \Big[
2\ \log\big({\textstyle \frac{E^\star}{\Lambda}}\big)+
\log\Big(1+\big({\textstyle \frac{h}{E^\star}}\big)^2\Big)\,\Big]\ , 
\eea
contains the   non-universal term  which
depends  on the UV cutoff scale $\Lambda$.
Using the Wick theorem, the
two-point functions \eqref{psoapaspas}  can be written 
in terms of
$D_\pm\equiv D(y|\pm h)$ \eqref{oaspasaso},
\bea\label{aopsosapssp}
&&\langle\,\mathfrak{M}_B(y)\,\mathfrak{M}_B(0)\,\rangle=m^2+4D_-D_+\\
&&\langle\,\mathfrak{S}_B(y)\,\mathfrak{S}_B(0)\,\rangle_{\rm conn}=
\frac{D_-D_+}{\mu^2}\,\bigg(\frac{{\ddot D}_-}{D_-}+
\frac{{\ddot D}_+}{D_+}-2\,\frac{{\dot D}_-{\dot D}_+}{D_-D_+}
+8h\Big(\frac{{\dot D}_+}{{ D}_+}-\frac{{\dot D}_-}{{ D}_-}\Big)+16h^2\bigg) .\nonumber
\eea
Here the dot stands for the derivative w.r.t.
the Euclidean time $y$ and
the abbreviation ``conn'' means the connected correlation
function:
$\langle\,\mathfrak{S}_B(y)\,\mathfrak{S}_B(0)\,\rangle_{\rm conn}
\equiv
\langle\,\mathfrak{S}_B(y)\,\mathfrak{S}_B(0)\,\rangle-
\langle\,\mathfrak{S}_B\,\rangle^2$. 
Similarly, one has
\bea\label{aopapssp}
\langle\,\mathfrak{S}_B(y)\,\mathfrak{M}_B(0)\,\rangle=
\frac{2 D_-D_+ }{\mu}\ \bigg(\frac{{\dot D}_+}{D_+}-\frac{{\dot D}_-}{D_-}-4h\,\bigg)
\ .
\eea

Since the two-point functions \eqref{aopsosapssp},\,\eqref{aopapssp}
are available in  closed forms,
the partition function
${\bar {\cal Z}}_{21}\big(\sigma_0|y\big)$
can be calculated perturbatively in powers of
$\delta m\equiv m_2-m_1\ll 1$ and $\delta\alpha\equiv
(E^\star_2-E^\star_1)/E^\star_1\ll 1$. 
In  the case $\delta m=m=0$,  details of the  calculations
may be found in Appendix in ref.\cite{Vasseur}.
Similar calculations for non-zero $ m$ and $\delta m$
yield  the first non-vanishing terms 
of the  Taylor expansion  of the 
fidelity $A_{21}(\sigma_0)$ in  the Toulouse limit:
\bea
A_{21}(\sigma_0)|_{g=\frac{1}{2}}=
\bigg(\frac{\pi\cos(\frac{\pi m}{2})}{4\re^{\gamma_E} E^\star}\bigg)^{d_{21}}\
\bigg(1-\chi_{\alpha\alpha}\ \frac{(\delta\alpha)^2}{2}-
\chi_{mm}\ \frac{(\delta m)^2}{2}+
O(\delta^3)\bigg)\ .
\eea
Here $m=m_1,\ E^\star=E_1$, 
$d_{21}=(\delta m)^2/16$ and
\bea
\chi_{\alpha\alpha}&=&
\frac{1}{4\pi^2}\ \bigg(1+{\frac{\pi}{2}}(m+1)\ \tan\Big(\frac{\pi m}{2}\Big)\bigg)
\bigg(1+{ \frac{\pi}{2}}(m-1)\ \tan\Big({ \frac{\pi m}{2}}
\Big)\bigg)\nonumber \\
\chi_{mm}&=&\frac{\pi^2\ \chi_{\alpha\alpha}}{\sin^2(\pi m)}-
\frac{\cos(\pi m)}{16\sin^2(\frac{\pi m}{2})}\ \bigg(1+
\pi m\, \tan\Big(\frac{\pi m}{2}\Big)\,\bigg)\ .
\eea

\section{\label{Sec4}Jost operators}

In this section we introduce the notion of 
quantum Jost operators and discuss their properties.

\subsection{\label{class}Classical  Jost functions}


Let us  first  consider the  limit $g\to 0$ where the 
boundary field $\Pi_B(t)={\dot \Phi}_B(t)$
is treated  as a classical field.
In this   approximation the Hamiltonian
 \eqref{islsasa}
describes the quantum spin
in the presence of a classical background field.
The corresponding non-stationary  Schr${\rm
{\ddot o}}$dinger equation
has the form
\bea\label{nonst}
\ri\, \frac{\partial \Psi}{\partial t} = -\big(\,  J\, \sigma_1
+ {\dot\phi}_c(t)\, \sigma_3\, \big)\ \Psi\ ,
\eea
where 
we use
$\phi_c(t)=\frac{1}{2}\, \Phi_B(t) + h t$,
satisfying   the asymptotic condition
\bea
\phi_c(t)\to h t+o(1)\ \ \ \ \ \ {\rm as}\ \ \  t\to-\infty
\eea
with $h>0$.
Let ${\boldsymbol U}(t,t_0)$ be a time evolution matrix for \eqref{nonst}.
It can be written as
\bea\label{lsasajks}
 {\boldsymbol U}(t,t_0)=\re^{\ri \sigma_3\phi_c(t)}
{\boldsymbol S}(t,t_0)
\ \re^{-\ri \sigma_
3\phi_c(t_0)}\ ,
\eea
where ${\boldsymbol  S}$ stands for a     time-ordered  matrix exponential 
\bea
{\boldsymbol  S}(t,t_0)
={\cal T}_t\exp\bigg(\ri\,
\int_{t_0}^{t} \rd t\, J\,
\big(\,
\re^{2\ri\phi_c(t)}\, \sigma_-+\re^{-2\ri\phi_c(t)}\, \sigma_+
\,\big)\, \bigg)\ .
\eea
Consider the time evolution of the spin-up state $|\uparrow\,\rangle$ 
starting from the initial moment $t_0$.
The  following limiting behavior for
$t_0\to-\infty$,  can be  easily established:
\bea\label{jasghshgs}
\re^{\ri\phi_c(t)\sigma_3} 
{\boldsymbol  S}(t,t_0)  \,  |\,\uparrow\,\rangle
&\to&
\bigg( \frac{{\tt k}+h}{ 2 {\tt k}}\ \Psi_{+{\tt k}}(t)\, 
|\,\uparrow\,\rangle+\frac{J}{ 2
{\tt k}}\ {\tilde \Psi}_{+{\tt k}}(t)\,
|\,\downarrow\,\rangle   \bigg)\ \re^{-\ri ({\tt k}-h) t_0} 
\\ 
&+& \bigg(\, \frac{{\tt k}-h}{ 2 {\tt k}}\ \Psi_{-{\tt k}}(t)\,
|\,\uparrow\,\rangle-\frac{J}{ 2
{\tt k}}\ {\tilde \Psi}_{-{\tt k}}(t)\,
 |\,\downarrow\,\rangle \, \bigg)\ \re^{\ri ({\tt k}+h) t_0}\ .\nonumber
\eea
Here 
$\Psi_{\pm {\tt k}}(t)$, ${\tilde \Psi}_{\pm {\tt k}}(t)$  
are   the Jost solutions
of the Sturm-Liouville equations
\bea\label{slasklsad}
&&\big(-\partial_t^2+U(t)\, \big)\ \Psi_{\pm {\tt k}}=J^2\, \Psi_{\pm {\tt k}}
\ ,\ \ \ \ \ U=- {\dot \phi}^2_c+
\ri \ {\ddot \phi}_c\\
&&\big(-\partial_t^2+
{\tilde U}(t)\, \big)\ {\tilde \Psi}_{\pm {\tt k}}=J^2\, {\tilde \Psi}_{\pm {\tt k}}
\ ,\ \ \ \ \ {\tilde U}
=- {\dot \phi}^2_c-
\ri\ {\ddot \phi}_c\ ,\nonumber
\eea
satisfying 
the   asymptotic conditions at $t\to-\infty$:
\bea\label{kslksa}
\Psi_{\pm{\tt k}}(t)\to \re^{\pm \ri {\tt k} t}\ ,\ \ \ \ 
{\tilde \Psi}_{\pm{\tt k}}(t)\to \re^{\pm \ri {\tt k} t}\ , 
\eea
where  ${\tt k}=\sqrt{J^2+h^2}>0$.
Because of  the presence of   oscillating phase factors,
the r.h.s.  of \eqref{jasghshgs} does not possess a  finite limit as $t_0$ tends to $-\infty$.
However,
if we  assume that the coupling $J$ is switched   off  adiabatically,
\bea\label{slslsa}
\lim_{t_0\to-\infty}J(t_0)= 0\ ,
\eea
then 
$\lim_{t_0\to-\infty}{\tt k }=h>0$ and  the first term in   \eqref{jasghshgs}  has a  finite limit.
The  second  term   will still  oscillate   $\propto \re^{2\ri h t_0}$ as  $t_0\to-\infty$.
With the  aim to suppress these oscillations, let us fix  the  value of $t$, say $t=0$, and
assume that the asymptotic behavior  \eqref{jasghshgs} holds true in  the
infinitesimal  wedge $0< \arg(-t_0)<\epsilon\to +0$
of   complex plane $t_0$. Then,  taking the limit along any  ray inside the wedge,
one obtains
\bea\label{lkklas}
{\rm lim}_{|t_0|\to +\infty\atop
\arg (-t_0)=+0}\ \Big(\, 
\re^{\ri\phi_c(0)\sigma_3}\,{\boldsymbol  S}(0,t_0)\, 
|\,\uparrow\,\rangle\, \Big)=\sqrt{\frac{1+m}{2}}\ \Big(\,
T^{(c)}_+\ |\,\uparrow\,\rangle+T^{(c)}_-\,|\,\downarrow\,\rangle\, \Big)\  ,
\eea
where we use  $m=\frac{h}{\sqrt{J^2+h^2}}$.  
The $t$-independent  connection  coefficients
\bea\label{soisao}
T^{(c)}_+=\sqrt{\frac{1+m}{2}}\ \ \Psi_{+{\tt k}}(0)\ ,\ \ \ \ \ 
T^{(c)}_-=\sqrt{\frac{ 1-m}{2}}
\ \ 
\ {\tilde \Psi}_{+{\tt k}}(0)
\eea
are sometimes referred as  to   Jost functions.
It is easy to see that eqs.\eqref{slasklsad},\,\eqref{kslksa} imply that 
\bea 
{\tilde \Psi}_{\pm\tt k}(t)=
\frac{\ri}{h\mp {\tt k}}\ \big(\partial_t-\ri{\dot\phi}_c\big) \Psi_{\pm \tt k}(t)\ ,\ \  \ 
{ \Psi}_{\pm\tt k}(t)=
-\frac{\ri}{h\pm {\tt k}}\ \big(\partial_t+\ri{\dot\phi}_c\big) {\tilde \Psi}_{\pm \tt k}(t)\nonumber
\eea
and also 
${\tilde \Psi}_{\mp\tt k}(t)=\Psi^*_{\pm \tt k}(t)$ for real $J,h$ and $t$.
Using this and also taking into account that 
the Wronskian of 
$\Psi_{- \tt k}(t)$ and $\Psi_{+ \tt k}(t)$  equals $2\ri {\tt k}$,
one finds the bilinear relation
\bea\label{osasopsasop}
\big(T^{(c)}_+\big)^*T^{(c)}_+ +\big(T^{(c)}_-\big)^*T^{(c)}_-=1\ .
\eea

Comparing  \eqref{soisao} with eqs.\eqref{osapas},\,\eqref{osapasa}
we note that within the mean field 
approximation, the classical connection coefficients
$\Psi_{+{\tt k}}(0)$  and ${\tilde \Psi}_{+{\tt k}}(0)$ are substituted by the
exponential operators  
$\re^{\frac{\ri m}{2}\Phi_B}(0)$ and $\re^{-\frac{\ri m}{2}\Phi_B}(0)$,
respectively.

\subsection{Anticipated properties of Jost operators}

Motivated by the above  consideration,
we start from the   Hamiltonian \eqref{aosaps} which describes
an  interaction of
a local spin impurity with a  chiral bose field on the  whole line.
Consider the interaction picture, treating the term  $\propto \mu$  as a perturbation.
Let us  perform the Wick rotation from the very beginning, 
so that $\phi(x,y)=\phi(x+{\rm i}y)$
for the chiral  bose field in the interaction picture and
introduce the $y$-ordered exponent
\bea
\label{Smatrix}
{\boldsymbol S}(y_2,y_1\,|\,\mu_L,h)=
{\cal T}_y\exp\left( \int^{y_2}_{y_1}{\rm d}y\  \mu_L(y)\,
\left({\rm e}^{2{\rm i}\phi(0,y)}\,{\rm e}^{2h y}\sigma_-+
{\rm e}^{-2{\rm i}\phi(0,y)}\,{\rm e}^{-2h y}\ \sigma_+\right)\,
\right)\ .
\eea
Here the renormalized  coupling  
$\mu$ is substituted   by $y$-dependent function, $ \mu_L(y)>0$
such that $\mu=\mu_L(0)$, and
which is switching off 
adiabatically within     an Euclidean time interval $|y|<L$.
For technical reason,  it is also  convenient to choose the 
smoothing function 
to be an even function of $y$.
The  explicit  IR regularization allows one to define
\begin{eqnarray}
\label{Lmatrix}
{\boldsymbol U}^{(-)}(\mu_L,h)&=&{\rm e}^{\ri\phi(0,0)\sigma_3}\, {\boldsymbol S}(0,-\infty\,|\,\mu_L,h)\, 
\\
{\boldsymbol U}^{(+)}(\mu_L,h)&=&
{\boldsymbol S}(+\infty,0\,|\,\mu_L,h)\, {\rm e}^{-\ri\phi(0,0)\sigma_3}\ ,\nonumber
\end{eqnarray}
which are  operators acting on the impurity spin, 
whose matrix elements are themselves operators 
acting on the free bosonic degrees of freedom, i.e. in the Hilbert space 
${\cal H}$ \eqref{integ}.
In the absence of interaction and for $h>0$,
the vacuum state 
is given by  the product 
$|\,{\rm vac}\,\rangle\otimes |\uparrow\,\rangle$.
Naively, the vacuum
in the interacting theory  occurs through the large-$L$ limit:
\bea\label{ooaspa}
\Big[\, \re^{\Delta f(L)}\ 
\ {\boldsymbol U}^{(-)}(\mu_L,h)\,
|\,{\rm vac}\,\rangle\otimes|\uparrow\,\rangle\,\Big]_{L\to+\infty}
&\to&|\,\Omega\,\rangle\\
\Big[\,\re^{\Delta f(L)}\ \ 
\langle\,{\rm vac}\,|\otimes\langle\uparrow|\,
{\boldsymbol U}^{(+)}(\mu_L,h)\,\Big]_{L\to+\infty}&\to&\langle\,
\Omega\,|\ ,
\nonumber
\eea
where $\Delta f(L)=\int_{-\infty}^0{\rm d} y\,
\big( E_0\big(\mu_L(y), h\big)+h\big)$.
But, because the IR problem,
these asymptotic  relations
cannot be literally true.
We will try to overcome this obstacle
using a  heuristic picture which is based
on the notion of
quantum Jost operators. 
Namely, we  postulate that
the exact  vacuum state is given by  relations
similar to the one obtained in  Sec.\,\ref{Sec3.1} within
the mean field approximation, i.e.,
%
\begin{eqnarray}
\label{oapasps}
|\,\Omega\,\rangle&=&
T_+(\mu,h)\, |\,{\rm vac}\,\rangle\otimes |\,\uparrow\,\rangle+T_-(\mu,h)\,
|\,{\rm vac}\,\rangle\otimes |\downarrow\,\rangle\\
\langle\, \Omega\,|&=&\langle\,{\rm vac}\,| \, Q_+(\mu,h)\otimes\langle\, 
\uparrow|+
\langle\,{\rm vac}\,| \, Q_-(\mu,h)\otimes\langle\, \downarrow|
\ ,\nonumber
\end{eqnarray}
and $T_\varepsilon$ and 
${Q}_\varepsilon$,  with  $\varepsilon=\pm$  and $\mu,h>0$, are
operators  acting as
\bea\label{aopsaasop}
&&T_\varepsilon (\mu,h)\ \ :\ \ \ 
{\cal H}\mapsto {\cal H}\ \ \&\ \ 
{\cal F}_p\mapsto {\cal F}_{p+\frac{m}{2}}\ \ \ \ \ \ \ \ \ 
\nonumber\\
&&Q_\varepsilon (\mu,h)\ \ :\ \ \
{\cal H}\mapsto {\cal H}\ \ \&\ \  {\cal F}_p\mapsto {\cal F}_{p-\frac{m}{2}}\ ,
\eea
such that  $p$-vacuum
expectation values  involving bilinear combinations of   $T$ and $Q$
are expressed in terms of the dimensionless amplitudes ${\cal A}_{\pm}$
\eqref{un}:
\bea\label{unaye}
\langle\,p'\,|\, Q_{\pm}(
\mu_2,h_2)\,
T_{\pm}(\mu_1,h_1)\,|\,p\,\rangle=
{\cal  A}_{\pm}\big( m_2,
\,m_1\,|\,\alpha\big)\ \delta_{2p'-2p,m_1-m_2}\ .
\eea
Notice that, 
since $T$ and $Q$
act invariantly in ${\cal H}$,
the $p$-dependence  
appears here  through
the Kronecker  delta only.
In what follows $T$- and $Q$-operators
are  referred as to   Jost operators.
In order to predict  their properties,
we shall invoke the  intuition which
is     based 
on the classical limit, 
the results of     perturbative calculations and 
 global  symmetries of the model.

\begin{itemize}


\item $\boldsymbol{ T-invariance.}$ 
The  time reversal transformation acts as
$\re^{\pm \ri\phi(0,y)}\mapsto \re^{\mp \ri\phi(0,-y)}$ and 
$\sigma_\pm\mapsto \sigma_\mp, \ \sigma_3\mapsto-\sigma_3$.
The  antiunitary  operator defined by
\eqref{CTRA}, satisfies the
relation
$${\mathbb T}\big(\re^{\ri a_1\phi_1(0,y_1)}\,
\re^{\ri a_2\phi(0,y_2)}\cdots \re^{\ri a_n\phi_1(0,y_n)}\big) {\mathbb T}=
\re^{\ri a_n\phi(0,-y_n)}\,\cdots\,
\re^{\ri a_{2}\phi(0,-y_{2})}\, \re^{-\ri a_1\phi_1(0,-y_1)}
$$
for $y_1>y_2>\cdots> y_n$. 
Ignoring  the problem with the IR  divergency, one can
expand the $y$-ordered exponent \eqref{Smatrix}  and find
\bea\label{psaaso}
{\mathbb T}\,  {T}_{\varepsilon} \big(\mu,h\big)\, {\mathbb T}=
{Q}_{\varepsilon} \big(\mu^*,h\big)\ ,\ \ \ \  \ \ \ 
{\mathbb T}\, {Q}_{\varepsilon} \big(\mu,h\big)\,  {\mathbb T}=
{T}_{\varepsilon} \big(\mu^*,h\big)\ .
\eea
Notice that the  $T$-invariance, when it is applied to 
\eqref{unaye},
leads  to the   relation \eqref{fornf}.

\item $\boldsymbol{Hermiticity.}$ 
The Jost operators satisfy a  hermitian conjugation condition
\bea\label{asosaoas}
Q_{\varepsilon}(\mu,h)=
\big(T_{\varepsilon}(\mu,h)\big)^\dagger\ \ \ \ \ \ \ \ (\mu,\,h>0)\ ,
\eea
which follows from the relations
$
\big({\rm e}^{{\rm i}\phi(0,0)\sigma_3}\big)^\dagger=
{\rm e}^{-{\rm i}\phi(0,0)\sigma_3}$,\ 
$\big({\rm e}^{2{\rm i}\phi(0,y)}\,\sigma_-\big)^\dagger=
{\rm e}^{-2{\rm i}\phi(0,-y)}\, \sigma_+$.
Together with \eqref{fornf}, the conjugation 
implies the  reality 
condition 
$\big({\cal A}_{\pm}(m_2,m_1|\alpha)\big)^*=
\ {\cal A}_{\pm}(m_2,m_1|\alpha)$
for real positive  $\mu_{i},\,h_{i} \ (i=1,2)$.

\item $\boldsymbol{ C-invariance.}$ 
The formal ${C}$-transformation 
acts as $\re^{2\ri a\phi(0,y)}\mapsto
\re^{-2\ri a\phi(0,y)}$, $\sigma_\pm \mapsto\sigma_\mp$,
$\sigma_3 \mapsto -\sigma_3$.
It is  well  known that
there is no
spontaneous magnetization in 
the anisotropic Kondo model.
Because 
of this 
we  expect that there exists an  operator ${\mathbb C}$ such that
${\mathbb C}^2=1$,
and the Jost  operators for $h<0$ can be
introduced through the relations 
\bea\label{asopsaaso}
{T}_{-\varepsilon}\big(\mu,-h\big)&=&
{\mathbb C}\, {T}_{\varepsilon} \big(\mu,h\big)\, {\mathbb C}\\
{Q}_{-\varepsilon}\big(\mu,-h\big)&=&
{\mathbb C}\, {Q}_{\varepsilon} \big(\mu,h\big)\, {\mathbb C}\ .\nonumber
\eea
An important consequence of the Hermiticity, $C$-  and  $ T$-invariance is that
 $p$-vacuum  expectation values  of   Jost operators
are expressed in terms of 
a single,  real analytic function  of $m\in (-1,1)$:\footnote{
Notice that as it follows from \eqref{soisao},
  $\lim_{g\to 0}F (m)=\big(\frac{1+ m}{2}\big)^{\frac{1}{2}}$.}
\bea
\label{uauasa}
\,\langle\, p'\,|\,
T_\pm  (\mu,h)\, |\,p\,\rangle
&=&
F(\pm m)\ \delta_{2p'-2p, m}
\\
\,\langle\, p'\,|\, Q_\pm (\mu,h)\, |\,p\,\rangle&=&
F(\pm m )
\ \delta_{2p-2p', - m}\ .
\nonumber
\eea
As for the matrix elements \eqref{unaye}, the
$C$-invariance
implies that ${\cal A}_{\pm}$ satisfy the relation \eqref{relsus}
within the principal  domain \eqref{phys}.


\item $\boldsymbol{Bilinear\ relation.}$
As it follows from eq.\eqref{normab}, ${\cal A}_{+}(m_2,m_1\,|0)+{\cal A}_{-}(m_2,m_1\,|0)=1$.
Taking this into account along with
the conjugation rule  \eqref{asosaoas}, the quantized version of the bilinear relation  \eqref{osasopsasop}
 is expected to be given by 
\bea\label{SMART}
\sum_{\varepsilon=\pm  }Q_\varepsilon(\mu,h) T_\varepsilon(\mu,h)={1}|_{\cal H}\ .
\eea

\item $\boldsymbol{ Lorentz\  invariance.}$ 
Let us introduce the complex coordinate $z=x+{\rm i}\, y $ and the polar angle $\psi=\arg({\rm i}\, z)$.
The generator of infinitesimal Euclidean rotations
coincides with $(-{\rm i}\, K)$, where  $K$ is
the Lorentz boost generator.
It is crucial for our analysis that
the  angular  evolution of Jost operators
turns out to  be remarkably  simple. Namely,
\bea
\label{oapsaspas}
{\rm e}^{-{\rm i} \psi K}\ T_\varepsilon(\mu,h)\ {\rm e}^{+{\rm i} \psi K}&=&
T_\varepsilon\big(\re^{\ri  (1-g)\psi}\mu, \re^{\ri\psi}h \big)\\
{\rm e}^{-{\rm i} \psi K}\ Q_\varepsilon(\mu,h)\ {\rm e}^{+{\rm i} \psi K}&=&
Q_\varepsilon\big(\re^{\ri  (1-g)\psi}\mu, \re^{\ri\psi}h \big)\,  .\nonumber
\eea
This follows from three observations;
First,  the  exponential fields
in \eqref{Smatrix}
are chiral (holomorphic)  $\big({\rm e}^{\pm 2{\rm i}\phi(x,y)}\equiv{\rm e}^{\pm 2\ri\phi}(z)\big)$
with the    Lorentz
spin $g$. Second,
the exponential operator located at the origin
is not affected by  the Euclidean rotation:
${\rm e}^{-{\rm i} \psi K}\ \re^{\pm \ri \phi(0,0)}\ {\rm e}^{+{\rm i} \psi K}=
\re^{\pm \ri \phi(0,0)}$.
Finally, 
we have to    accept that
integrals containing
combinations of  the {\it holomorphic} fields are not  changed by
rotations of the integration contour in the limit  $L\to\infty$.

The simple  geometry of the Euclidean plane
suggests  that  $T$- and $Q$-  operators  are related  
through 
the Euclidean rotation   of angle $\pi$, combined with  the $C$-conjugation:
\bea
\label{conj}
Q_\varepsilon(\mu,h)
={\mathbb C}\, {\rm e}^{-\ri\pi  K}
\ T_{\varepsilon}(\mu,h)\ {\rm e}^{\ri\pi  K}\, {\mathbb C}\ ,
\eea
or, equivalently  (see eqs.\eqref{asopsaaso},\eqref{oapsaspas})
\bea
\label{conja}
Q_\varepsilon(\mu,h)= T_{-\varepsilon}\big({\rm e}^{\ri\pi(1-g)}\mu,h\big)\ .
\eea
It is worth to keep in mind 
that these  formulae
should  be understood in a weak sense
as relations   for   the analytic continuation
of a certain class  of   matrix elements of Jost operators.
The applicability of eqs.\eqref{oapsaspas} 
(which is crucial for deriving \eqref{conja}) 
requires that 
the   contours of integration,   involving  
in the  construction   of Jost operators,
can be  rotate  freely   within the Euclidean plane.
This may be  not the case  for   general matrix elements.

We  expect that \eqref{conja} can be applied 
for the matrix elements \eqref{unaye}, and therefore  the dimensionless  amplitudes
$ {\cal  A}_{\pm}$
are  related by   the analytic 
continuation
with  two-point functions containing
$T$-operators only:
\bea\label{unayw}
\langle\,p'\,|\, T_{\mp}(
\mu_2,h_2)\,
T_{\pm}(\mu_1,h_1)\,|\,p\rangle=
{\cal  A}_{\pm}\big(\re^{-\ri\pi}\, m_2,
\,m_1\,|\,\alpha-\ri\pi\big)\ 
 \delta_{2p'-2p,m_1+m_2}\ .
\eea
Here the  phase rotation $\re^{-\ri\pi}$ means
the analytic continuation
along the clockwise  half-circle  of radius smaller then one 
(see Fig.\,\ref{fig4})  
for
$-1<m_1<1$ and $\alpha\in {\mathbb R}$.
\begin{figure}
\centering
\psfrag{+1}{$+1$}
\psfrag{-1}{$-1$}
\psfrag{m}{$m_2$}
\includegraphics[width=6cm]{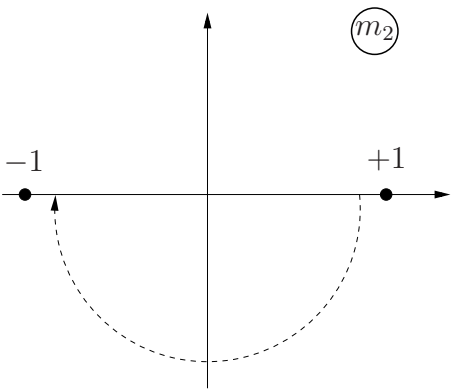}
\caption{The contour of analytic continuation in eq.\eqref{unayw} as  
$m_1\in (-1,1)$ and $\alpha\in {\mathbb R}$.}
\label{fig4}
\end{figure}
Notice  that the analytic 
continuation appearing in the r.h.s. of
\eqref{unayw}
does not involve any problem
for the perturbative amplitudes \eqref{aspass},\,\eqref{relsus}.\footnote{
In this connection
it deserves  mentioning
that eq.\eqref{relsus}   should be understood  as
${\cal A}_{\varepsilon}(m_2,m_1\,|\,\alpha)=
{\cal A}_{\varepsilon}(              
\re^{\pm \ri\pi} m_2, \re^{\pm \ri\pi}m_1\,|\,\alpha)$,
which   is  satisfied for  any  sign $\pm$.}
This  supports
rather sweeping
assumptions that have been
made  in  derivation of \eqref{unayw}.

\end{itemize}

\section{\label{Sec5}Algebra of  Jost operators}


At the moment it is not clear how to deal with
the fidelities
beyond the scope of perturbation theory  
for arbitrary values
$\mu_{i}$ and $h_i$. However, in the case of $h_1=h_2\equiv h$  
much can be said about the  matrix elements
\bea\label{unsusyw}
\langle\,p'\,|\, T_{\varepsilon_2}
(\mu_2,h)\,
T_{\varepsilon_1}(\mu_1,h)\,|\,p\,\rangle=F_{\varepsilon_2 \varepsilon_1}(m_2, m_1)\ 
\ \delta_{2p'-2p, m_2+m_1}\ .
\eea
Notice that according to eq.\eqref{unayw} 
\bea\label{aoisa}
{\cal  A}_{\pm}\big( m_2,
\,m_1\,|\,\alpha\big)\big|_{h_1=h_2}=F_{\mp\pm}(\re^{\ri\pi}m_2,m_1)\ ,
\eea
and   $\alpha$ 
is not
an independent variable   as  $h_1=h_2$. 
In this case $\alpha$ can be 
written as
\bea
\alpha=
\frac{1}{1-g}\  \log\bigg(\frac{\lambda_2}{\lambda_1}\bigg)\ ,
\eea
where $\lambda_i\ (i=1,2)$ stands for the dimensionless ratio $\mu_i/h^{1-g}$,
which is a certain function of $m_i$, i.e., $\lambda_i=\lambda(m_i)$.
The corresponding inverse function $m=m(\lambda)$ is given by
eq.\eqref{ioaiso}.
In the present discussion we will   use both  variables   $m_i$ and 
$\lambda_i$ assuming that  they  are related through 
the formula \eqref{ioaiso}.
Since $h>0$ is   assumed  to be fixed from now on,
it makes sense to simplify the
notation for the  Jost operators:
\bea
T_{\varepsilon}
(\mu,h)\equiv T_{\varepsilon}(m)\ ,\ \ \ \ Q_{\varepsilon}
(\mu,h)\equiv Q_{\varepsilon}(m)\ .
\eea

\subsection{Commutation relations}

Let us recall that  $\phi(x,y)$ at $y=0$ satisfies   the  commutation relation \eqref{com}.
Since the Euclidean time dependence of the chiral field $\phi(x,y)$    is  very simple,
we can 
translate  \eqref{com}
into the commutation relation  at   $x=0$:
\bea\label{coma}
[\phi(0,y_2),\,\phi(0,y_1)]=\frac{\ri}{2}\ \pi g\ \sgn(y_2-y_1)\ .
\eea
According to ref.\cite{BLZ}, if  the matrix-valued operators
${\boldsymbol L}_i= \re^{\ri\phi(0,y_2)} {\boldsymbol S}(y_2,y_1\,|\,\mu_i,h)$
$(i=1,2)$,
where 
${\boldsymbol S}$ is  defined  by eq.\eqref{Smatrix} with  $y_2>y_1$ and
with $\mu_i$     are set to be $y$-independent constants,
then the Yang-Baxter equation is satisfied in the form
\bea
{\check {\boldsymbol R}}\,\big(\,{\boldsymbol L}_1\otimes 1\,\big)\,
\big(\,1\otimes {\boldsymbol L}_2\,\big)=
\big(\,1\otimes {\boldsymbol L}_2\,\big)\, 
\big(\,{\boldsymbol L}_1\otimes 1\,\big)\,{\check {\boldsymbol R}}\, . \label{YB}
\eea
Nontrivial matrix 
elements of the  $4\times 4$ matrix ${\check {\boldsymbol R}}$ read explicitly
 \begin{eqnarray}
 {\check R}_{++}^{++}={\check R}_{--}^{--}=1\,,\ \ \ \ \ \ \ \
  {\check R}_{+-}^{+-}={\check R}_{-+}^{-+}=\frac{\lambda_1^2-\lambda_2^2}{{\tt q}\lambda_1^2-{\tt q}^{-1}\lambda_2^2}\, ,
 \ \ \ \ 
    {\check R}_{+-}^{-+}={\check R}_{-+}^{+-}=\frac{ ({\tt q}-{\tt q}^{-1})\lambda_1\lambda_2}
{{\tt q}\lambda_1^2-{\tt q}^{-1}\lambda_2^2} 
\,,
    \end{eqnarray}
where we  use  ${\tt q}=\re^{\ri\pi g}$ and $\lambda_i=\mu_i/h^{1-g}$.
Although a mathematically satisfactory  construction
of the Jost operators is absent,
 the arguments similar to those from
ref.\cite{Sergei1} suggest   
that the Jost operators  satisfy  the    commutation relations
\bea
\label{ZF}
T_{\varepsilon_1}(m_1)
T_{\varepsilon_2}(m_2)=\sum_{\varepsilon_1',\varepsilon_2'=\pm}
R_{\varepsilon_1\varepsilon_2}^{\varepsilon_1'\varepsilon_2'}(m_1,m_2)\ 
T_{\varepsilon'_2}(m_2)T_{\varepsilon'_1}(m_1)\ ,
\eea
where
the $R$-matrix
 obeys   the Yang-Baxter equation  together with  
the  ``unitarity'' and   ``crossing symmetry'' relations
\bea
&&\sum_{\varepsilon_1',\varepsilon_2'=\pm}
R_{\varepsilon_1\varepsilon_2}^{\varepsilon_1'\varepsilon_2'}(m_1,m_2)
R_{\varepsilon'_2\varepsilon'_1}^{\varepsilon_3\varepsilon_4}(m_2,m_1)=
\delta_{\varepsilon_1}^{\varepsilon_4}
\delta_{\varepsilon_2}^{\varepsilon_3}\ ,\\
&&\sum_{\varepsilon_1',\varepsilon_2',\varepsilon_3'=\pm}\delta^{\varepsilon'_1+\varepsilon_3',0}\,
R_{\varepsilon'_1\varepsilon'_2}^{\varepsilon_1\varepsilon_2}(\re^{\ri\pi}m_1,m_2)\,
R_{\varepsilon'_3\varepsilon_4}^{\varepsilon_3\varepsilon'_2}(m_1,m_2)=
\delta^{\varepsilon_1+\varepsilon_3,0}
\, \delta^{\varepsilon_2}_{\varepsilon_4}\ .
\nonumber
\eea
The formal r${\hat{\rm o}}$le of 
the  Yang-Baxter  and   unitarity relations is clear;
the  unitarity is required for  self-consistency of \eqref{ZF},
whereas the Yang-Baxter equation is  the associativity constraint. 
In its turn, the crossing symmetry allows one to supplement 
 algebraic relations  \eqref{ZF}  with 
an  extra  bilinear relation,
\bea\label{SMA}
\sum_{\varepsilon=\pm  }T_{-\varepsilon}(\re^{\ri\pi} m)\, 
T_\varepsilon(m)= 1\ ,
\eea
which follows from eqs.\,\eqref{SMART} and \eqref{conja}.

The  $R$-matrix  in the  form
\bea\label{aopsosap}
R_{\varepsilon_1\varepsilon_2}^{\varepsilon_1'\varepsilon_2'}(m_1,m_2)=
R(m_1,m_2)\
 {\check R}_{\varepsilon_1\varepsilon_2}^{\varepsilon_1'\varepsilon_2'}\ ,
\eea
meet all the necessary  requirements, if
 the normalization factor  satisfies   the conditions
\bea\label{sapsa1}
&&R(m_1,m_2)\, R(m_2,m_1)=1\\
&&R(m_1,m_2)\, R(\re^{\ri\pi} m_1,m_2)=\frac{{\tt q}\lambda_1^2-{\tt q}^{-1}\lambda_2^2}{\lambda_1^2-\lambda_2^2}\ .
\nonumber
\eea
(recall that we use the convention  $\lambda_i=\lambda(m_i)$).
Some extra  conditions are imposed by the global symmetries. Namely, the
Hermiticity and $T$-invariance require that
\bea\label{sapsa2}
R(m_1,m_2) R^*(m_1,m_2)=1\ \ \ \ \ {\rm for}\ \ \ 0<m_{1,2}<1\ ,
\eea
whereas the 
 $C$-symmetry yields the relation
\bea\label{sapsa3}
R\big( \re^{\pm\ri\pi}m_1, \re^{\pm \ri\pi}m_2\big)=R(m_1,m_2)\ .
\eea

\subsection{Normalization factor $R(m_1,m_2)$}

The algebra of the Jost operators    together with the global
symmetry relations and normalizations conditions lead to   a system  of functional 
equations  imposed  on  the  two-point function
$F_{\varepsilon_2\varepsilon_1}(m_2,m_1)$  \eqref{unsusyw}.
Assuming that   $0<m_{1,2}<1$,
the system reads as follows:
\bea\label{aopsoa1}
F_{\varepsilon_1\varepsilon_2}(m_1,m_2)=\sum_{\varepsilon_1',\varepsilon_2'=\pm} 
R_{\varepsilon_1\varepsilon_2}^{\varepsilon_1'\varepsilon_2'}(m_1,m_2)
F_{\varepsilon'_2\varepsilon'_1}(m_2,m_1)
\eea
and
\bea\label{aopsoa2}
F_{\varepsilon_2\varepsilon_1}(\re^{\ri\pi}m_2, m_1)=F_{-\varepsilon_1,-\varepsilon_2}( \re^{\ri\pi}m_1,m_2)=
F_{-\varepsilon_2,-\varepsilon_1}( m_2,\re^{-\ri\pi}m_1)=
\big(F_{\varepsilon_2\varepsilon_1}(\re^{\ri\pi}m_2,  m_1)\big)^*
\eea
and
\bea\label{aopsoa}
F_{-\varepsilon \varepsilon }(\re^{\ri\pi}m,m\,|\,0\,)=
\frac{1}{2}\ (1+\varepsilon\,   m)
\ .
\eea
Using the perturbative results \eqref{aspass},
one can check that  all the   conditions for $F_{\mp\pm}$  are  satisfied
at the first perturbative order, provided
the $R$-matrix has the form \eqref{aopsosap} 
with
\bea\label{aopsopsap}
\log R(m_1,m_2)=\frac{\ri}{2}\,  \pi g\ \ \frac{m_2-m_1}{m_2+m_1}\ \ (1+m_2 m_1)+O(g^2)\ .
\eea
Without additional analytical assumptions the set 
\eqref{sapsa1}-\eqref{sapsa3} and \eqref{aopsopsap} 
does  not unambiguously define $R(m_1,m_2)$.
However, the first perturbative correction
allows one to make a conjecture  about  the exact  normalization factor.
Namely, a  simple calculation shows that, at least at the first perturbative 
order, the normalization factor obeys   the condition
\bea\label{main}
\bigg(\,\lambda_2\,\frac{\partial}{\partial \lambda_2}+
\lambda_1\,\frac{\partial}{\partial \lambda_1}\,\bigg)\ 
\log R(m_1, m_2)=\frac{\ri}{2}\  \pi g\
\bigg(\,\lambda_2\,\frac{\partial}{\partial \lambda_2}-
\lambda_1\,\frac{\partial}{\partial \lambda_1}\,\bigg)\ m_1 m_2\ .
\eea
Here $m_i\equiv
m(\lambda_i)$\ $(i=1,2)$.
If we accept   \eqref{main}  for any
$0<g<1$,
the reconstruction  of $\log R(m_1,m_2)$ requires 
a minimum
amount of additional  analytical
assumptions. Indeed, consider the limit when  $h\to 0$, keeping the Kondo temperatures
$E^\star_1$ and $E^\star_2$ fixed. 
In this case   both $m_1$ and $m_2$ turns to be zero, but their
ratio remains finite. Then
\bea\label{aoipsa}
\lim_{\mathfrak{h}\to 0}R\big( \mathfrak{h}
\re^{\alpha_1}, \mathfrak{h}\re^{\alpha_2}\big)=R_0(\alpha_1-\alpha_2)\ ,
\eea
where $\alpha_1-\alpha_2= \log(E^\star_2/E^\star_1)$, 
and  as it follows from eqs.\eqref{sapsa1},
\bea\label{hasu}
R_0(\alpha)R_0(-\alpha)=1\ ,\ \ \ \ R_0(\alpha+\ri\pi) 
R_0(\alpha)=-\frac{\sinh(1-g)(\alpha+\ri\pi)}{\sinh (1-g)\alpha}\ .
\eea
(Notice  
that for unrelated  $h_1$ and $h_2$ the variable $\alpha$
coincides with the one   defined by eq.\eqref{gamma}.)
The ``minimal'' solution of \eqref{hasu}  
(i.e., such that  $\log R_0(\alpha)$ is 
analytic in the strip $0\leq \Im m(\alpha)\leq \pi $ and  bounded
at $\alpha\to+\infty$) has the form
\bea\label{aisosia}
R_0(\alpha)=\exp\left(-{\rm i}\,\int_{0}^\infty 
\frac{{\rm d} \omega}{ \omega}\ 
\frac{\sin(\alpha\omega )}
{\cosh \frac{\pi\omega}{2}}\ \frac{\sinh \frac{\pi g\omega}{2(1-g)}}
{\sinh \frac{ \pi\omega}{2(1-g)}}\,
\right)\, .
\eea
Eq.\eqref{main}, as a linear partial differential equation subject
of the asymptotic condition \eqref{aoipsa}, 
can be easily solved by means of  the Fourier transform. 
Using \eqref{ioaiso} one finds
\bea\label{asakjas}
R(m_1,m_2)=\exp\bigg(
\frac{g}{ 8\pi\ri}
{\iint}_{-\infty}^{+\infty}\,
\rd\omega_1\rd\omega_2\ \lambda_1^{\frac{\ri\omega_1}{1-g}}
\lambda_2^{\frac{\ri\omega_2}{1-g}}
\ \  \frac{M(\omega_1)\, M(\omega_2)}
{ (\omega_1+\ri 0)(\omega_2+\ri 0)}\   \frac{\omega_2-\omega_1}{ \omega_2+\omega_1-\ri 0}\bigg)\,  .
\eea

The function \eqref{asakjas} meets    
the required  conditions \eqref{sapsa1}-\eqref{sapsa3}. It also
possesses  a small-$g$ series expansion
\bea\label{aslaskl}
\log R(m_1,m_2)=\ri\pi\, \sum_{n=1}^{\infty}g^n\ r_n(m_1,m_2)\ ,
\eea
with
\bea\label{slksl}
&&r_1=\frac{1}{ 2}\ \frac{m_2-m_1}{ m_1+m_2}\ (1+m_1\ m_2)\nonumber \\
&&r_2=\frac{m_1 m_2 (1-m_1^2) (1-m_2^2)}{(m_1^2-m_2^2)^2}\ \bigg(\,
m_1^2- m_2^2+2m_1 m_2\log\Big(\frac{m_2}{m_1}\Big)\,\bigg)\\
&&r_3=\frac{m_1 m_2 (1-m_1^2) (1-m_2^2)}{(m_1^2-m_2^2)^3}\ \bigg(\,
(m_1^2-m_2^2)^2\ (1+m_1^2+m_2^2-3 m_1
m_2)-\nonumber\\ &&
\frac{\pi^2}{ 12}\ (m_1-m_2)^4\ (m_1^2+m_2^2+
m_1 m_2-1)+2\ m_1
m_2(2m^2_1\,m^2_2-m^2_1-m^2_2)\
\log^2\Big(\frac{m_2}{ m_1}\Big)\,\bigg)\nonumber,etc.
\eea
It would be valuable to confirm  these  higher-order corrections
using the renormalized perturbation theory.
More importantly, the  origin of  \eqref{main} (if it is true, of course)
needs to be clarified.
It may be useful to note here  that the relation \eqref{main} 
can be equivalently  written as
\bea
\Big(\,\lambda_2\,\frac{\partial}{\partial \lambda_2}+
\lambda_1\,\frac{\partial}{\partial \lambda_1}\,\Big)\
\log {\tilde R}(m_1, m_2)=\ri\pi\ 
\Big(\,\lambda_2\,\frac{\partial}{\partial \lambda_2}-
\lambda_1\,\frac{\partial}{\partial \lambda_1}\,\Big)\ d_{21}(-m_2,m_1)\ ,
\eea
where ${\tilde R}(m_1, m_2)=\re^{\frac{\ri\pi g}{4}(m_2^2-m_1^2)}\ R(m_1, m_2)$ 
and $d_{21}=
 \frac{g}{4}\, (m_2-m_1)^2$ is the orthogonality exponent.


\subsection{\label{Sec5.4}Fidelities  $A_{21}(\sigma_{0,3})$
for $h=0$}

As $h\to 0$, the system of functional equations \eqref{aopsoa1}-\eqref{aopsoa}  is simplified dramatically.
In this case
\bea
\lim_{\mathfrak{h}\to 0^+}   F_{\pm\mp}
\big( \mathfrak{h} \re^{\alpha_2}, \mathfrak{h} \re^{\alpha_1} )=f(\alpha_1-\alpha_2)
\eea
and  $f$  satisfy the equations:
\bea\label{apaop}
f(\ri\pi+\alpha)=
f(\ri\pi-\alpha)\ ,\ \ \ \frac{f(-\alpha)}{f(\alpha)}=\frac{\sinh\frac{1-g}{2}\,(\ri\pi+\alpha)}
{\sinh\frac{1-g}{2}\, (\ri\pi-\alpha)}\ R_0(\alpha)\ .
\eea
Let us assume now that $\log f$ 
is  analytic
in the strip $0<\Im  m(\alpha)\leq \pi$  and grows slower then any
exponential of $\alpha$ at infinity.\footnote{
These assumptions  can be supported by the perturbative 
results \eqref{aspass}. Also note
that as  $g=0$ and $h=0$,
the natural domain of analyticity of the classical  Jost functions  $T_\pm^{(c)}$ 
\eqref{soisao},
as  functions of the complex variable  
 $J={\tt  k}|_{h=0}$, is the lower  half plane $-\pi\leq \arg ({  J})\leq 0$
which corresponds to the strip  $0\leq \Im m (\alpha)\leq\pi$.}
With these  analytical assumptions,  the  functional relations   \eqref{apaop}
define $f$   up to a constant multiplicative factor.
Then the  normalization condition \eqref{aopsoa} 
fixes $f$   unambiguously.
Finally, using relations \eqref{aoisa}, we derive
\bea\label{aiosapsa}
A_{21}(\sigma_0)|_{h_1=h_2=0}&=&2\,f(\alpha-\ri\pi)=u(\alpha)
\ 
\frac{\sinh (1-g)\,\frac{\alpha}{2}}{ (1-g) \sinh\frac{\alpha}{ 2}}\ 
\nonumber\\
A_{21}(\sigma_3)|_{h_1=h_2=0}&=&0\ ,
\eea
where
\bea\label{gxi}
u(\alpha)=\exp\left(\int_0^\infty \frac{{\rm d}\omega}{  \omega} 
\frac{\sin^2 (\alpha\omega/2 )}{
\sinh \pi\omega \cosh\frac{\pi\omega}{2} }\ \ 
\frac{\sinh \frac{\pi g\omega }{2(1-g)}}{
\sinh \frac{\pi\omega}{2(1-g)}}\right)\, .
\eea
A numerical verification  of the prediction \eqref{aiosapsa}
was  performed  in ref.\cite{Vasseur}.

\section{Concluding remarks}

Although the problem of finding  fidelities $A_{21}(\sigma_s)$ 
for
non-vanishing $h$ remains unsolved, it seems useful to
discuss its generalization.

The boundary operators $\sigma_s$ 
are particular representatives of  the 
family of
 ``primary'' boundary  fields
\bea
\sigma_s^{(a)}= \re^{\ri a\Phi_B}\ \sigma_s\  \ \ \ \ \ \ (s=0,3,\pm)
\eea
(here $\sigma_\pm=\sigma_1\pm \ri\sigma_2$).
Recall that the Hamiltonian ${\boldsymbol H}$\eqref{islsasa}
relates to the renormalized Hamiltonian 
${\boldsymbol H}_\phi$ \eqref{aosaps}
by means of the canonical
transformation
${\boldsymbol  H}_\phi= 
{\boldsymbol U}^\dagger {\boldsymbol H} {\boldsymbol  U}$
 with ${\boldsymbol U}=\exp\big(\ri\sigma_3\phi(0)\big)$
and $\Phi_B(0)=2\phi(0)$. 
For this reason
the  scale dimension of   
$\sigma_\pm\re^{\ri a\Phi_B}$ is given by $D_\pm(a)=g (a\mp 1)^2$ whereas
the scale dimension of   
$\sigma_{0,3}\re^{\ri a\Phi_B}$ equals  to $D_{0,3}(a)=g a^2$.
An overall normalization of the  primary boundary fields
can be chosen to satisfy
the normalization condition at $y\to 0^+$,
\bea\label{isuyjasuasi}
&&\sigma_s^{(-a)}(y)\sigma_s^{(a)}(0)\to  \sigma_0\
y^{-2 ga^2}+\ldots  \ \ \ \ \ \ \ \ \ \ \ \ \ \ \ \ \ \ \  \
\ \ \ \ \ (s=0,3)\nonumber\\
&&\sigma_{-s}^{(-a)}(y)\sigma_{s}^{(a)}(0)\to  \frac{1}{2}\ (\sigma_0-s\, \sigma_3)\
y^{-2 g(a-s)^2}+\ldots  \ \ \ \ \ \ \ \  (s=\pm)\  ,
\eea
along with the hermitian conjugation  
\bea
\big[\sigma_s^{(a)}(y)\big]^\dagger=
\begin{cases}
\sigma_s^{(-a)}(-y)\ \ \  \ \ \ &{\rm for}\ \ \ s=0,3\\
\sigma_{-s}^{(-a)}(-y)\ \ \  \ \ &{\rm for}\ \ \ s=\pm 
\end{cases}\ .
\eea

Repeating the construction from Section\,\ref{Sec2.3}, one can
introduce
the normalized partition function
${\bar {\cal Z}}_{21}\big({\cal O}|y\big)$ \eqref{aopasopsapas}
for the    operators ${\cal O}=\sigma_s^{(a)}$.
As $y\to 0^+$, 
the effect of  the  boundary interaction
becomes negligible    within
the interval $(0,y)$ 
and
the  pair of hermitian conjugated operators  
at the ends of the segment can be substituted by
a local insertion
defined by the leading term of  the operator product expansions
\eqref{isuyjasuasi}.
This yields  the leading  behavior at $y\to 0^+$:
\bea
{\bar {\cal Z}}_{21}\big(\sigma_s^{(a)}|y)
=\begin{cases}
  y^{-2 g a^2 }\, \big(1+o(1)\big)\ \ \ \ &{\rm for}\ \ \
s=0,3\\
\frac{1}{2}\, (1-s m_1)\ y^{-2 g (a-s)^2 }\, \big(1+o(1)\big)\ \ \ \ &{\rm for}\
\ \
s=\pm
\end{cases}\ .
\eea
The spin degrees of freedom are freezing out at large distances so that
the large-$y$ asymptotic of
${\bar {\cal Z}}_{21}\big({\sigma}^{(a)}_{s}|y\big)$
has a  similar   form as   eq.\eqref{kssus},
\bea\label{kssussewr}
{\bar {\cal Z}}_{21}\big({\sigma}^{(a)}_{s}|y\big)
=
\big|A_{21}\big({\sigma}^{(a)}_{s}\big)\big|^2\ \
y^{-2  d_{21}(a)}\
   \re^{-y \Delta E_{21}}\ \big(1+o(1)\big)\ \ \ \
\ \ {\rm as}\ \ \ y\to+\infty\ .
\eea
The IR  exponent $d_{21}(a)$ can be found  using  the
mean field approximation  from  Section\,\ref{Sec3.1}
\bea\label{expoa}
d_{21}(a)=\frac{g}{4}\ (m_2-m_1-2a)^2\ .
\eea
Just as in the case $a=0$, this formula is expected to be
an exact result. This  supposition is supported by a  perturbative
calculation 
given in the appendix.
As before,
 $A_{21}\big(\sigma_s^{(a)})$ in \eqref{kssussewr} can be thought
as
scaling  functions depending on
the     $m_1$, $m_2$, and the pair of
Kondo temperatures $E^\star_1$, $E^\star_2$.
In the context of the spin-boson model \eqref{lsasa},
the fidelity
$A_{21}\big(\sigma_s^{(a)})$  represents the 
universal part of the vacuum-vacuum  matrix element
$\langle\,\Omega_2\,|\,\big[\sigma_s^{(a)}\big]_{\rm bare}\,|
\,\Omega_1\rangle$
of the ``bare'' operator \eqref{aopoas}, which
requires both UV and IR regularizations.
If the integration
in  \eqref{lsasa} and \eqref{aopoas}
is restricted to the finite  
domain   $L^{-1}<k<\Lambda$, then
\bea
\langle\,\Omega_2\,|\,\big[\sigma_s^{(a)}\big]_{\rm bare}\,|
\,\Omega_1\rangle\propto  A_{21}\big(\sigma_s^{(a)}\big)\ 
L^{-d_{21}(a)}\ \Lambda^{-D_s(a)}\ .
\eea

In the case $h_1=h_2$,
the calculation of       $A_{21}\big(\sigma_s^{(a)}\big)$
can be  reduced to 
finding   $p$-vacuum  matrix  elements
generalizing \eqref{unsusyw}:
\bea\label{unsyw}
\langle\,p'\,|\, \re^{2\ri a \phi(0)}\, T_{\varepsilon_2}
(m_2)\,
T_{\varepsilon_1}(m_1)\,|\,p\,\rangle=
F^{(a)}_{\varepsilon_2 \varepsilon_1}(m_2, m_1)\
\ \delta_{2p'-2p,2a+ m_2+m_1}\ .
\eea
The bosonization approach developed  in ref.\cite{Sergei1} allows one
to study  the case $h_1=h_2\to 0^+$ and
extend the result from  Section\,\ref{Sec5.4}.
It turns out that,
for the primary field $\sigma^{(a)}_+$ with $0\leq a\leq 1$,
the fidelity
is simply expressed in terms of
the function $u(\alpha)$   \eqref{gxi}:
\bea\label{majsj}
A_{21}\big(\sigma^{(a)}_+\big)\big|_{h_1=h_2\to 0^+}=
{ {\cal S}}_+(a)\ \ u(\alpha)
\ \big(E^\star_1E^\star_2\big)^{ g (\frac{1}{2}- a)}\  ,
\eea
where $\alpha=\log(E^\star_2/E_1^\star)$. 
In the case of $\sigma^{(a)}_{0,3}$, the corresponding fidelities 
are given by
\bea\label{majsja}
A_{21}\big(\sigma^{(a)}_s\big)\big|_{h_1=h_2\to 0^+}=
{\cal S}_0(a)\ \big(\,\re^{\ri\pi ga}\
{\cal A}^{(a )}(\alpha)+(-1)^s\
\re^{-\ri\pi ga}\  {\cal A}^{( -a )}(\alpha)\,\big)\ \ \ \ \ 
(s=0,3)\ ,
\eea
where
\bea\label{kaoasio}
{\cal A}^{(a)}(\alpha)&=&
u(\alpha)\
\int_{-\infty}^{+\infty}\frac{\rd\gamma}{2\pi}\ \frac{\re^{ga (2\gamma-\alpha)}}{\cosh\gamma}\
\frac{u(\gamma+\frac{\ri\pi}{2}) u(\gamma-\frac{\ri\pi}{2}) }{
u(\gamma-\alpha+\frac{\ri\pi}{2}) u(\gamma-\alpha-\frac{\ri\pi}{2})}\ .
\eea
Note that ${\cal S}_+(a)$ and ${\cal S}_0(a)$ in  
eqs.\eqref{majsj},\,\eqref{majsja}
stand for
some dimensionless 
functions of $a$ and the coupling $g$, which remained  undetermined
except their values at $a=0$:
\bea
{  {\cal S}}_+(0)
=-\frac{1}{\pi g }\ \tan\Big(\frac{\pi g}{2(1-g)}\Big)\ \ \Gamma(1-g)
\Bigg(\frac{g\sqrt{\pi}\Gamma(\frac{g}{2(1-g)})}
{2\Gamma(\frac{1}{2}+\frac{g}{2(1-g)})}\Bigg)^{1-g}\ ,\ \ \ \ 
S_0(0)=1\ .
\eea

Finally,
one can   consider the
fidelities for  scaling fields of the form 
$$P_{N}\big(\partial_y\Phi_B,\partial^2_y\Phi_B,\ldots\big)\ 
\re^{\ri a\Phi_B}\   \sigma_s\ ,$$
where $P_{N}(\ldots)$ stands for a differential polynomial of 
order $N$ built from the boundary field $\partial_y\Phi_B$.
In this most general setting,  
the problem is   related   to finding 
the
multipoint amplitudes
\bea\label{unsywyst}
\langle\,p'\,|\, \re^{2\ri a \phi(0)}\, T_{\varepsilon_n}
(m_n)\,\cdots
T_{\varepsilon_1}(m_1)\,|\,p\,\rangle=
F^{(a)}_{\varepsilon_n\ldots \varepsilon_1}(m_n,\ldots, m_1)\
\delta_{2p'-2p,-2a+ m_n+\cdots+m_1}\ .
\eea  

\section*{Acknowledgments} 

\noindent
I would like to thank N. Andrei,  F.A. Smirnov and A.B. Zamolodchikov for useful discussions,
and especially H. Saleur for many valuable  discussions, comments on the manuscript and encouragement.
This work was supported by the National 
Science Foundation Grant No.1404056.

\appendix

\section{\label{CPT}Appendix:
Conformal Perturbation Theory}

Here we  consider the large $y$ asymptotic of the  
normalized partition function  
${\bar {\cal Z}}_{21}\big(\sigma^{(a)}_{0}|y\big)$
for $E^\star_1=0,\ h_1>0$, or equivalently,  $m_1=1$.
Note that in this case
\bea
{\bar {\cal Z}}_{21}\big({\sigma}^{(a)}_{3}|y\big)=
{\bar {\cal Z}}_{21}\big(\sigma^{(a)}_{0}|y\big)\ ,\ \ \ \ \ \ \
{\bar {\cal Z}}_{21}\big({\sigma}^{(a)}_{-}|y\big)=0\ .
\eea

\begin{figure}
\centering
\psfrag{a}{$a$}
\psfrag{-a}{$-a$}
\psfrag{+1}{$+1$}
\psfrag{-1}{$-1$}
\psfrag{h}{$h$}
\includegraphics[width=14  cm]{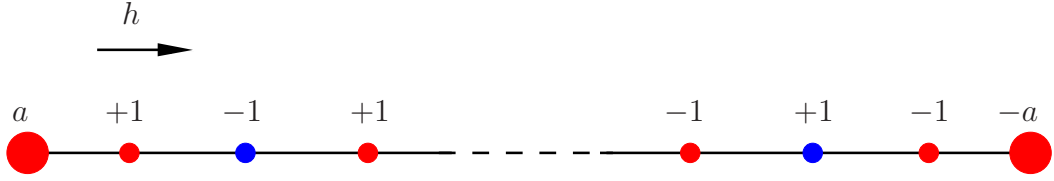}                      
\caption{The Anderson-Yuval \cite{AndersonYuval}  one-dimensional 
gas 
 with an external  field $h$ and external   charges 
$\pm a$ at  the ends of segment.
The arrow shows   the  direction of 
the  external force exerted on the ``positive'' charges.} 
\label{fig3}
\end{figure}

\begin{figure}
\centering
\psfrag{A}{$(E^\star=0,\ h>0)$
}
\psfrag{B}{$(E^\star>0,\ h=0^+)$}
\psfrag{C}{$\re^{-\ri a \Phi_B}$}
\psfrag{CFT}{Gaussian\  CFT}
\psfrag{L}{$L\to\infty$}
\includegraphics[width=4cm]{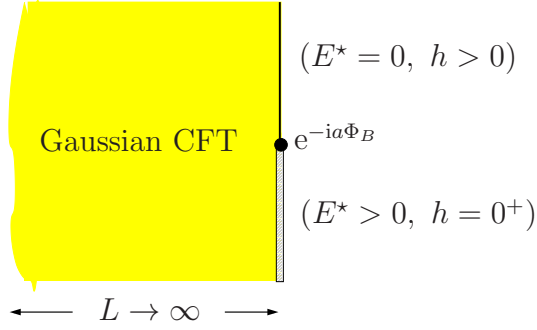}
\caption{The dimensionless amplitude
${\cal A}_0^{(a)}$ \eqref{kajsak}  is the  universal part of the
partition function of the statistical system 
schematically shown in this figure.}
\label{figI4}
\end{figure}

Using the renormalized Hamiltonian \eqref{aosaps},
where $\mu$ is renormalized coupling corresponding to  
the Kondo temperature $E^\star\equiv E^\star_2$ and
$h\equiv h_2$, the ratio   
${\bar {\cal Z}}_{21}\big(\sigma^{(a)}_{0}|y\big)=
{ {\cal Z}}_{21}\big({\sigma }^{(a)}_{0}|y\big)/{{\cal Z}_{11}}$
 can be represented in the form of the grand canonical
partition function  of  the Anderson-Yuval  one-dimensional gas
of alternating charges on the finite interval,
\bea\label{lslsys}
{\bar {\cal Z}}_{21}(\sigma^{(a)}_{0}|y)
\big|_{E^\star_1=0}&= &\sum_{n=0}^\infty\mu^{2n}\      
\int_{0}^{y}\rd y_{2n}\int_{0}^{y_{2n}}\rd y_{2n-1}
\cdots\int_{0}^{y_{2}}\rd y_{1}\prod_{l=1}^{n}\re^{2h(y_{2l-1}-y_{2l})} \\
&\times&\langle\, p\,|\, \re^{-2\ri a \phi (\tau)}\
\re^{-2\ri \phi (y_{2n})} \re^{+2\ri \phi (y_{2n-1})} \cdots
\re^{-2\ri \phi (y_{2})}\, \re^{2\ri \phi (y_{1})}\, \re^{+2\ri a \phi (0)} |\, p\,\rangle\ ,\nonumber
\eea
where, in fact, 
the matrix element does not depend  on the choice of  $p$-vacuum,
\bea\label{kaassaasui}
\big\langle\, p\,\big|\,
\re^{2\ri a_n\phi }(y_{n})\cdots
\re^{2\ri a_1\phi }(y_{1})\,\big|\, p\,\big\rangle=
\delta_{a_1+\cdots+a_n,0}\ \prod_{i>j}
(y_i-y_j)^{ 2g\, a_i a_j}\ .
\eea
The parameter $h$ plays the r${\hat {\rm o}}$le of the external electric field
whereas the
exponentials $\re^{\pm 2\ri a\phi }$
can be interpreted as  creators of  two external charges
at the ends of the interval (see Fig.\,\ref{fig3}).
Then one has
\bea\label{slksalsa}              
\big|A(\sigma^{(a)}_{0})\big|^2_{E^\star_1=0} 
=h^{2ga^2-2{\tilde  d}_{21}}\,
\lim_{{\ell }=hy\to+\infty} {\ell}^{2{\tilde  d}_{21}-2ga^2}
\re^{{\ell} (1-e_0(\lambda))}\ \bigg(\, 1+ 
\sum_{n=0}^{\infty}\lambda^{2n}
\ q_{ n}(\ell)\,\bigg)\ .
\eea
Here ${\tilde d}_{21}$ is some (a priori  unknown)
orthogonality exponent,
\bea\label{asopsaoaspas}
e_0(\lambda)\equiv -E_0/h>0\ ,\ \ \ \ \ \lambda\equiv \mu/h^{1-g}\ ,
\eea
and
\bea\label{jasasias}
q_{n}({\ell})&=&{\ell}^{2 n(1-g)}
\int_{0}^{1}\rd u_{2n}\int_{0}^{u_{2n}}\rd u_{2n-1}
\cdots\int_{0}^{ u_{2}}\rd u_{1}\
\prod_{i=1}^n\re^{2{\ell} (u_{2i-1}-u_{2i})}\
\bigg[\frac{(1-u_{2i}) u_{2i-1}}{ 
(1-u_{2i-1}) u_{2i}}\bigg]^{2 a g}\nonumber\\
&\times &\prod_{i,j=1}^{n}|u_{2i}-u_{2j-1}|^{-2 g}\
\prod_{1\leq i<j\leq n}\big|(u_{2j}-u_{2i})
(u_{2j-1}-u_{2i-1})\big|^{2g}\ .
\eea
Consider the first nontrivial coefficient $q_1({\ell})$.
One can  show  that as ${\ell} \to+\infty$,
\bea\label{sopsaas}
q_1({\ell})=
2^{ 2 g-1}\,  \Gamma(1-2 g)\ {\ell}
-
2^{ 2 g-2}\, \Gamma(2-2 g)\ \Big(\, 1+4a g\ \log\big( {\ell}\re^{C}\big)\, \Big)+
o(1)\ ,
\eea
where 
$C=1+\log(2)-\gamma_E-\psi(2-2 g)-\psi(1+2a  g)$ and  $\psi(z)\equiv\partial_z\log \Gamma(z)$.
Substituting  \eqref{sopsaas}
in \eqref{slksalsa} we observe that
the  term  $\propto \lambda^2{\ell}$  is canceled out provided that
$e_0(\lambda)=1+
2^{2g-1}\,  \Gamma(1-2 g)\ \lambda^2
+O(\lambda^4)$.
Taking into account that 
$m(\lambda)= e_0-(1-g)\lambda\,\frac{\partial e_0}{\partial \lambda}$, i.e.,
\bea
m(\lambda)= 
1- 2^{2g-1}\, \Gamma(2-2 g)\ \lambda^2
+O(\lambda^4)\ ,
\eea
one finds that the    
 cancellation of the  divergent term $\propto \lambda^2\log {\ell}$ 
requires that
\bea\label{oaaosap}
{\tilde   d}_{21}=g a^2+g a\, (1-m)+O\big((1-m)^2\big)\ ,
\eea
which is in agreement with \eqref{expoa} specialized for $m_1=1$, $m_2=m$.
The above   calculation  suggests that 
\bea\label{fin}
\big|A_{21}(\sigma^{(a)}_{0})\big|_{E^\star_1=0}
=h^{-ga (1-m)-\frac{g}{4}(1-m)^2}\ { A}(m,ag)\ ,
\eea
where
\bea
{ A}(m,l)=\sqrt{ \frac{1+m}{2}}\
\Big(\,1+\sum_{n=1}^\infty {\mathfrak a}_n(l)\ \ (1-m)^{n}\,\Big)\ ,
\eea
and the first expansion  coefficient reads explicitly
\bea\label{coeff}
{\mathfrak a}_1(l)=-l\ \big(\,
 1+\log(2)-\gamma_E-\psi(2-2 g)-\psi(1+2l )\, \big)\ .
\eea
The function \eqref{fin} is expected to have a finite limit as $h\to 0^+$:
\bea\label{kajsak}
\lim_{h\to 0^+}
\big|A_{21}(\sigma^{(a)}_{0})\big|_{E^\star_1=0}
=\big(E^\star\big)^{-g(a+\frac{1}{4})}\ {\cal A}_0^{(a)}\ .
\eea
The dimensionless amplitude 
${\cal A}_0^{(a)}$  is the  universal part of the 
partition function ${\cal Z}^{(a)}$ 
of the system depicted in Fig.\,\ref{figI4},
\bea\label{isoisoa}
{\cal Z}^{(a)}/{\cal Z}_{\rm free}\sim  
{\cal A}_0^{(a)}\ (c_1 \varepsilon E^\star )^{a^2} 
\  (c_2 LE^\star )^{-g(a+\frac{1}{2})^2}\ .
\eea
Here 
${\cal Z}_{\rm free}$  is the partition function
of the Gaussian CFT with free BC, $\varepsilon$ is the lattice spacing,  and  
$a$-independent,  non-universal coefficients $c_i$
depend on  details of the IR and UV regularizations.

It deserves to be mentioned here that
$\big|A_{21}(\sigma^{(a)}_{0})\big|_{E^\star_1=0}$
has    a finite limit
as $g\to 0$ and $l=ag$ is kept fixed.
In this case, as it follows from \eqref{lslsys},\,\eqref{kaassaasui},
\bea
\lim_{g\to 0\atop l=a g -{\rm fixed}}
y^{2ga^2 }\, {\bar {\cal Z}}_{21}\big(\sigma^{(a)}_{0}|y\big)\big|_{J_1=0}
=
\langle\,\uparrow|
{\cal T}_\tau\exp\bigg[J
\int_{0}^{y} \rd \tau 
\big(
\re^{2\ri\phi_c(-\ri \tau)}\, \sigma_-+\re^{-2\ri\phi_c(-\ri \tau)}\, \sigma_+
\,\big) \bigg]
|\uparrow\,\rangle
\eea
where
$\ri\, \phi_c(-\ri \tau)= h \tau+ l\ \log\big(\frac{\tau}{y-\tau}\big).$
In the presence of  singularities at $\tau=0,\,y $, 
the analysis performed in Section\,\ref{class} requires some modification.
One can show that ${ A}(m,l)$ at $g=0$ coincides with the connection coefficient
\bea\label{apaspoasosspa}
T^{(c)}_+(m,l)=
\sqrt{ \frac{1+m}{2}}\ 
\lim_{\tau\to 0^-} (-\tau)^{l}\, \Psi_{\tt k}(-\ri \tau)\ ,
\eea
where  $\Psi_{{\tt k}}(-\ri \tau)$
is the  Jost solution of
the Sturm-Liouville equation
\bea\label{laiua}
-\partial_\tau^2\Psi
+\Big(1-\frac{ 2{\tt k}l m}{  \tau}+\frac{l(l+1)}{\tau^2}
\, \Big)\ \Psi=0\ ,
\eea
satisfying the
asymptotic condition
\bea
\Psi_{{\tt k}}(-\ri \tau)\to (- \tau)^{- l m}\ 
\re^{{\tt k}\tau } \ \  \ \ \ \ {\rm as}\ \ \ \ \ \tau\to-\infty
\eea
(recall that we use the notations
 ${\tt k}=\sqrt{J^2+h^2}>0$ and  $m=h/{\tt k}$).
The  Jost solution  is expressed in terms of
the  confluent hypergeometric function:
\bea
\Psi_{{\tt k}}(-\ri \tau)
={\tt k}^{lm}\  (-{\tt k} \tau)^{1+ l}\ \re^{{\tt k}\tau}\ \ 
2^{1+l (1+m)}\ U\big(1+ (1+  m)l \, ,\, 2(1+ l)\,,\, -2{\tt k}\tau\big)\ .
\eea
This yields the result 
\bea\label{hsyst}
T^{(c)}_+(m,l)=
\sqrt{\frac{1+m}{2}}\ \   {m}^{l(1-m) }\ \  \frac{2^{l(m-1)}\Gamma(1+2l)}{
\Gamma(1+(1+m)l)}\ .
\eea
Notice that an    expansion of 
$T^{(c)}_+(m,l)$ at  $m=1$ turns out  to be consistent with the
classical limit of \eqref{fin}-\eqref{coeff}.
Thus we see that 
$\big|A_{21}(\sigma^{(a)}_{0})\big|_{E^\star_1=0\atop g=0}=
h^{-l}\, T^{(c)}_+(m,l)$, 
and
it  is an  analytic function for $\Re e (l)>-\frac{1}{2}$.
There are two other points which deserve notice, such as
the bilinear  relation 
\bea
T^{(c)}_+(\re^{\ri\pi} m,l)\,T^{(c)}_-(m,l)+
T^{(c)}_-(\re^{\ri\pi} m,l)\,T^{(c)}_+(m,l)=1\ ,
\eea
where 
\bea
T^{(c)}_-(m,l)=\sqrt{\frac{1-m}{2}}\ \    m ^{-l(1+m)}\    \ 
\frac{2^{l(m+1)}\Gamma(1-2l)}{
\Gamma(1-(1-m) l)}\ ,
\eea
(to be compared with eq.\eqref{SMA}),
and   the  behavior  of $T^{(c)}_+(l,m)$ as $m\to 0^+$:
\bea\label{ospaoasp}
T^{(c)}_+(m,l)\to {m}^{l }\ 2^l\ \frac{\Gamma(\frac{1}{2}+l)}{\sqrt{2\pi}}\ .
\eea

\end{document}